\documentclass[11pt,  draftclsnofoot,onecolumn,article,romanappendices]{IEEEtran}
\usepackage{setspace}

\usepackage{cite}
\ifCLASSINFOpdf
   \usepackage[pdftex]{graphicx}
\else
  \usepackage[dvips]{graphicx}
\fi
\usepackage{amsmath,amssymb,amsthm,mathrsfs,amsfonts,dsfont,stmaryrd}
\usepackage{array}
\usepackage{dsfont}
\usepackage{amsbsy}
\usepackage{epstopdf}
\usepackage{graphicx,color}
\usepackage{hyperref}

\usepackage{mathtools}

\usepackage{MnSymbol}

\long\def\symbolfootnote[#1]#2{\begingroup%
\def\thefootnote{\fnsymbol{footnote}}\footnote[#1]{#2}\endgroup}

\newtheorem{new-theorem}{Theorem}
\newtheorem{new-lemma}{Lemma}  
\newtheorem{new-corollary}{Corollary}  
\newtheorem{new-proposition}{Proposition}   
\newtheorem{new-definition}{Definition} 
\newtheorem{new-remark}{Remark} 
\newtheorem{new-claim}{\textbf{Claim}}

\newcommand{\dv}{\mathbf} 
\newcommand{\mc}{\mathcal} 
\newcommand{\mkv}{-\!\!\!\!\minuso\!\!\!\!-}

 \allowdisplaybreaks

\usepackage{algorithm}
\usepackage{algpseudocode}

\algnewcommand{\Inputs}[1]{%
  \State \textbf{Inputs:}
  \Statex \hspace*{\algorithmicindent}\parbox[t]{.8\linewidth}{\raggedright #1}
}
\algnewcommand{\Initialize}[1]{%
  \State \textbf{initialization}
  \Statex \hspace*{\algorithmicindent}\parbox[t]{.95\linewidth}{\raggedright #1}
}

\begin{document}

\title{On the Information Bottleneck Problems: Models, Connections,
Applications and Information Theoretic Views }
\author{Abdellatif Zaidi   \qquad \quad  I\~naki Estella Aguerri    \quad \quad Shlomo Shamai (Shitz)\\
\thanks{ Abdellatif Zaidi is with Universit\'e Paris-Est, France, and currently on leave at the Mathematical and Algorithmic Sciences Laboratory, Huawei Paris Research Center, 92100 Boulogne-Billancourt, France. I\~naki Estella Aguerri is with Telef\'onica Innovation Alpha, 08019, Barcelona, Spain.  Shlomo Shamai (Shitz) is with the Technion Institute of Technology, Technion City, Haifa 32000, Israel. The work of S. Shamai was supported by the European Union's Horizon 2020 Research And Innovation Programme, grant agreement No. 694630, and by the WIN consortium via the Israel minister of economy and science.

 Emails. \{abdellatif.zaidi@u-pem.fr, inaki.estella@telefonica.com,  sshlomo@ee.technion.ac.il\}. }}

\maketitle

\begin{abstract}
This tutorial paper focuses on the variants of the bottleneck problem taking an information theoretic perspective and discusses practical methods to solve it, as well as its connection to coding and learning aspects. The intimate connections of this setting to remote source-coding under logarithmic loss distortion measure, information combining, common reconstruction, the Wyner--Ahlswede--Korner problem, the efficiency of investment information, as well as, generalization, variational inference, representation learning, autoencoders, and others are highlighted. We discuss its extension to the distributed information bottleneck problem with emphasis on the Gaussian model and highlight the basic connections to the uplink Cloud Radio Access Networks (CRAN) with oblivious processing. For this model, the optimal trade-offs between relevance (i.e., information) and complexity (i.e., rates) in the discrete and vector Gaussian frameworks is determined. In the concluding outlook, some interesting problems are mentioned such as the characterization of the optimal inputs (``features'') distributions under power limitations maximizing the ``relevance'' for the Gaussian information bottleneck, under ``complexity'' constraints.
\end{abstract}


\IEEEpeerreviewmaketitle


\section{Introduction}

A growing body of works focuses on developing learning rules and algorithms using information theoretic approaches  (e.g., see \cite{TPB99, P10, YP18, YJP18, KS18, UE-AZ17a} and references therein). Most relevant to this paper is the Information Bottleneck (IB) method of~\cite{TPB99}, which seeks the right balance between data fit and generalization by using the mutual information as both   a cost function and a regularizer. Specifically, IB formulates the problem of extracting the relevant information that some signal $X \in \mathcal{X}$ provides about another one $Y \in \mathcal{Y}$ that is of interest as that of finding a representation $U$ that is maximally informative about $Y$ (i.e., large mutual information $I(U;Y)$) while being minimally informative about~$X$ (i.e., small mutual information $I(U;X)$). In the IB framework, $I(U;Y)$ is referred to as the {\textit{relevance}} of $U$ and $I(U;X)$ is referred to as the \textit{complexity} of $U$, where  complexity here is measured by the minimum description length (or rate) at which the observation is compressed. Accordingly, the performance of learning with the IB method and the optimal mapping of the data are found by solving the Lagrangian formulation 
\begin{equation}
\mc L^{\mathrm{IB},*}_{\beta}:=\max_{P_{U|X}} I(U;Y)-\beta I(U;X),
\label{eq:IBCriteria}
\end{equation}
where $P_{U|X}$ is a stochastic map that assigns the observation $X$ to a representation $U$ from which $Y$ is inferred and $\beta$ is the Lagrange multiplier. Several methods, which we detail below, have been proposed to obtain solutions $P_{U|X}$ to the IB problem in Equation         \eqref{eq:IBCriteria} in several scenarios, e.g., when the distribution of the sources $(X,Y)$ is perfectly known or only samples from it are available.

The IB approach,   as a method to both characterize performance limits as well as to design mapping, has found remarkable applications in supervised and unsupervised learning problems such as classification, clustering, and prediction. 
Perhaps key to the analysis and theoretical development of the IB method is its elegant connection with information-theoretic rate-distortion problems, as it is now well known that the IB problem is essentially a remote source coding problem~\cite{DT62,WW75,W80} in which the distortion is measured under logarithmic loss. Recent works show that this connection turns out to be useful for a better understanding the fundamental limits of learning problems, 
including the performance of deep neural networks (DNN)~\cite{S-ZT17}, the emergence of invariance and disentanglement in DNN~\cite{AS17}, the minimization of PAC-Bayesian bounds on the test error~\cite{M13, AS17}, prediction \cite{alemi2019variational, say2019machine}, or as a generalization of the evidence lower bound (ELBO) used to train variational auto-encoders~\cite{KW13, mukherjee2019general}, geometric clustering \cite{strouse2019information}, or extracting the Gaussian "part" of a signal \cite{PT18}, among others. Other connections that are more intriguing exist also with seemingly unrelated problems such as privacy and hypothesis testing~\cite{KC16, TC08, SGC18} or multiterminal networks with oblivious relays~\cite{E-AZCS17a,E-AZCS19a} and non-binary LDPC code design~\cite{stark2019decoding}. More connections with other coding problems such as the problems of information combining and common reconstruction, the Wyner--Ahlswede--Korner problem, and the efficiency of investment information  are unveiled and discussed in this tutorial paper, together with extensions to the distributed setting. 

The abstract viewpoint of IB also seems   instrumental to a better understanding of the so-called \textit{representation learning}~\cite{BCV13}, which is an active research area in machine learning that focuses on identifying and disentangling the underlying explanatory factors that are hidden in the observed data in an attempt to render learning algorithms less dependent on feature engineering.  More specifically, one important question, which is often controversial in statistical learning theory, is the choice of a ``good'' loss function that measures discrepancies between the true values and their estimated fits. There is however numerical evidence that models that are trained to maximize mutual information, or equivalently minimize the error's entropy, often outperform ones that are trained using other criteria such as mean-square error (MSE) and higher-order statistics~\cite{E02, PEL00}. On this aspect, we also mention Fisher's dissertation~\cite{F97}, which contains investigation of the application of information theoretic metrics to blind source separation and subspace projection using Renyi's entropy as well as what appears to be the first usage of the now popular Parzen windowing estimator of information densities in the context of learning. Although a complete and rigorous justification of the usage of mutual information as cost function in learning is still awaited, recently, a partial explanation appeared in~\cite{JCVW15}, where the authors showed that under some natural data processing property Shannon's mutual information uniquely quantifies the reduction of prediction risk due to side information. Along the same line of work, Painsky and Wornell~\cite{PW18} showed that, for binary classification problems, by minimizing the logarithmic-loss (log-loss), one actually minimizes an upper bound to any choice of loss function that is smooth, proper (i.e., unbiased and Fisher consistent), and convex. Perhaps, this justifies partially why mutual information (or, equivalently, the corresponding loss function, which is the log-loss fidelity measure) is widely used in learning theory and has already been adopted in many algorithms in practice such as the \textit{infomax} criterion~\cite{L88}, the tree-based algorithm of~\cite{Q14}, or the well known Chow--Liu algorithm~\cite{CL68} for learning tree graphical models, with various applications in genetics~\cite{AMPB08}, image processing~\cite{PMV03}, computer vision~\cite{VWW97}, etc. The logarithmic loss measure also plays a central role in the theory of prediction~\cite[Ch. 09]{C-BL06} where it is often referred to as the \textit{self-information} loss function, as well as in Bayesian modeling~\cite{LC06} where priors are usually designed     to   maximize the mutual information between the parameter to be estimated and the observations. The goal of learning, however, is not merely to learn model parameters accurately for previously seen data. Rather, in essence, it is the ability to successfully apply rules that are extracted from previously seen data to characterize new unseen data. This is often captured through the notion of ``generalization error". The generalization capability of a learning algorithm hinges on how sensitive   the output of the algorithm is to modifications of the input dataset, i.e., its \textit{stability}~\cite{BE02, SSSS10}. In the context of deep learning, it can be seen as a measure of how much the algorithm overfits the model parameters to the seen data. In fact, efficient algorithms should strike a good balance between their ability to fit training dataset and that to generalize well to unseen data. In statistical learning theory~\cite{C-BL06}, such a dilemma is reflected through that the minimization of the ``population risk" (or ``test error" in the deep learning literature) amounts to the minimization of the sum of the two terms that are generally difficult to minimize simultaneously, the ``empirical risk" on the training data and the generalization error.    To prevent over-fitting, regularization methods can be employed, which include parameter penalization, noise injection, and averaging over multiple models trained with distinct sample sets. Although it is not yet very well understood how to optimally control model complexity, recent works~\cite{XR17,RZ15} show that the generalization error can be upper-bounded using the mutual information between the input dataset and the output of the algorithm. This result actually formalizes the intuition that the less information a learning algorithm extracts from the input dataset the less it is likely to overfit, and justifies, partly,  the use of mutual information also as a regularizer term. {{The interested} reader may refer to \cite{Amjad_2019} where it is shown that regularizing with mutual information alone does not always capture all desirable properties of a latent representation}. We also point out that there exists an extensive literature on building optimal estimators of information quantities (e.g. entropy, mutual information), as well as their Matlab/Python implementations, including in the high-dimensional regime (see, e.g.,~\cite{P03,JVYW15,VV13,CMT16,AFDM17,AS18} and references therein). 

This paper provides a review of the information bottleneck method, its classical solutions, and recent advances. In addition, we unveil some useful connections with coding problems such as remote source-coding, information combining, common reconstruction, the Wyner--Ahlswede--Korner problem, the efficiency of investment information, CEO source coding under logarithmic-loss distortion measure, and learning problems such as inference, generalization, and representation learning. Leveraging  these connections, we discuss its extension to the distributed information bottleneck problem with emphasis on its solutions and the Gaussian model and highlight the basic connections to the uplink Cloud Radio Access Networks (CRAN) with oblivious processing. For this model, the optimal trade-offs between relevance and complexity in the discrete and vector Gaussian frameworks are determined. In the concluding outlook, some interesting problems are mentioned such as the characterization of the optimal inputs distributions under power limitations maximizing the ``relevance''  under ``complexity'' constraints.

\subsection*{Notation}

Throughout, uppercase letters  denote random variables, e.g., $X$;  lowercase letters denote realizations of random variables, e.g., $x$; and calligraphic letters denote sets, e.g., $\mathcal{X}$. The cardinality of a set is denoted by $|\mc X|$. For a random variable $X$ with probability mass function (pmf) $P_{X}$, we use $P_{X}(x)=p(x)$, $x\in \mc X$ for short. Boldface uppercase letters denote vectors or matrices, e.g., $\dv X$, where context should make the distinction clear. For random variables $(X_1,X_2,\hdots)$ and a set of integers $\mc K\subseteq \mathds{N}$,  $X_{\mc K}$ denotes the set of random variables with indices in the set $\mc K$, i.e., $X_{\mc K}=\{X_k:k \in \mc K \}$. If $\mc K = \emptyset$, $X_{\mc K}=\emptyset$. For $k\in \mc K$, we let $X_{\mathcal{K}/k} = (X_1,\hdots,X_{k-1},X_{k+1},\hdots,X_K)$, and assume that $X_0=X_{K+1}=\emptyset$. In addition, for zero-mean random vectors $\dv X$ and $\dv Y$, the quantities $\mathbf{\Sigma}_{\mathbf{x}}$, $\mathbf{\Sigma}_{\mathbf{x},\mathbf{y}}$ and $\mathbf{\Sigma}_{\mathbf{x}|\mathbf{y}}$  denote, respectively, the covariance matrix of the vector $\dv X$, the covariance matrix of vector $(\dv X,\dv Y)$, and the conditional covariance matrix of $\dv X$, conditionally on $\dv Y$, i.e., $\mathbf{\Sigma}_{\mathbf{x}} = \mathrm{E}[\mathbf{XX}^H]$ $\mathbf{\Sigma}_{\mathbf{x},\mathbf{y}}:= \mathrm{E}[\mathbf{XY}^H]$, and  $\mathbf{\Sigma}_{\mathbf{x}|\mathbf{y}} = \mathbf{\Sigma}_{\mathbf{x}}-\mathbf{\Sigma}_{\mathbf{x},\mathbf{y}}\mathbf{\Sigma}_{\mathbf{y}}^{-1}\mathbf{\Sigma}_{\mathbf{y},\mathbf{x}}$. Finally, for two probability measures $P_X$ and $Q_X$ on the random variable $X \in \mc X$, the relative entropy or Kullback--Leibler divergence is denoted as $D_{\mathrm{KL}}(P_X \| Q_X)$. That is, if $P_X$ is absolutely continuous with respect to $Q_X$, $P_X \ll Q_X$ (i.e., for every $x \in \mc X$, if $P_X(x) >0$, then $Q_X(x) >0 )$, $D_{\mathrm{KL}}(P_X \| Q_X) = \mathbb{E}_{P_X}[\log(P_X(X)/Q_X(X))]$, otherwise $D_{\mathrm{KL}}(P_X \| Q_X)=\infty$.

\section{The Information Bottleneck Problem}

\begin{figure}[!t]
\centering
\includegraphics[width=0.8\textwidth]{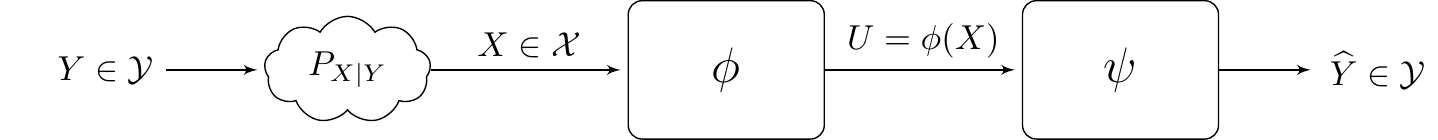}
\caption{Information bottleneck problem.} 
\label{fig-abstract-IB-model}
\end{figure}

The Information Bottleneck (IB) method was introduced by Tishby {et    al.}~\cite{TPB99} as a method for extracting the information that some variable $X \in \mathcal{X}$ provides about another one $Y \in \mathcal{Y}$ that is of interest, as shown in Figure~\ref{fig-abstract-IB-model}.

 Specifically, the IB method consists of finding the stochastic mapping $P_{U|X}:\mathcal{X}\rightarrow\mathcal{U}$ that from an observation $X$ outputs a representation $U\in\mc U$ that is maximally informative about $Y$, i.e., large mutual information $I(U;Y)$, while being minimally informative about $X$, i.e., small mutual information $I(U;X)$~({As such,} 
the usage of Shannon's mutual information seems to be motivated by the intuition that such a measure provides a natural quantitative approach to the questions of meaning, relevance, and common-information, rather than the solution of a well-posed information-theoretic problem---a connection with source coding under logarithmic loss measure appeared later on in~\cite{HT07}.) The auxiliary random variable $U$ satisfies that $U \mkv X \mkv Y$ is a Markov Chain in this order; that is, that the joint distribution of $(X,U,Y)$ satisfies
\begin{equation}
p(x,u,y)  = p(x)p(y|x)p(u|x), \label{MK-chain}
\end{equation}
and the mapping $P_{U|X}$   is chosen such  that $U$ strikes a suitable balance between the degree of \textit{relevance} of the representation as measured by the mutual information $I(U;Y)$ and its degree of \textit{complexity} as measured by the mutual information $I(U;X)$. In particular, such $U$, or effectively the mapping $P_{U|X}$, can be determined     to   maximize the IB-Lagrangian defined as 
\begin{equation}
\mc L_{\beta}^{\mathrm{IB}}(P_{U|X}):= I(U;Y)- \beta I(U;X) 
\label{IB-Lagrangian-formulation-minimizing-complexity}
\end{equation}
over all mappings $P_{U|X}$ that satisfy $U \mkv X \mkv Y$ and the  trade-off parameter $\beta$ is a positive Lagrange multiplier associated with the constraint on $I(U;Y)$. 

Accordingly, for a given $\beta$ and source distribution $P_{X,Y}$, the optimal mapping of the data, denoted by $P_{U|X}^{*,\beta}$,  is found by solving the IB problem, defined as
\begin{equation}
\mc L^{\mathrm{IB},*}_{\beta}:=\max_{P_{U|X}} I(U;Y)-\beta I(U;X).
\label{eq:IBCriteria},
\end{equation}
over all mappings $P_{U|Y}$ that satisfy $U \mkv X \mkv Y$. It follows from the classical application of Carath\'eodory's theorem \cite{GK11} that without loss of optimality, $U$ can be restricted to satisfy  $|\mc U| \leq |\mc X|+1$.

In Section \ref{sec:SolutionsIB} we discuss several methods to obtain solutions $P^{*,\beta}_{U|X}$ to the IB problem in Equation         \eqref{eq:IBCriteria} in several scenarios, e.g., when the distribution of  $(X,Y)$ is perfectly known or only samples from it are available.

\subsection{The IB Relevance--Complexity Region} 

\begin{figure}[!t]
\centering
\includegraphics[width=0.55\textwidth]{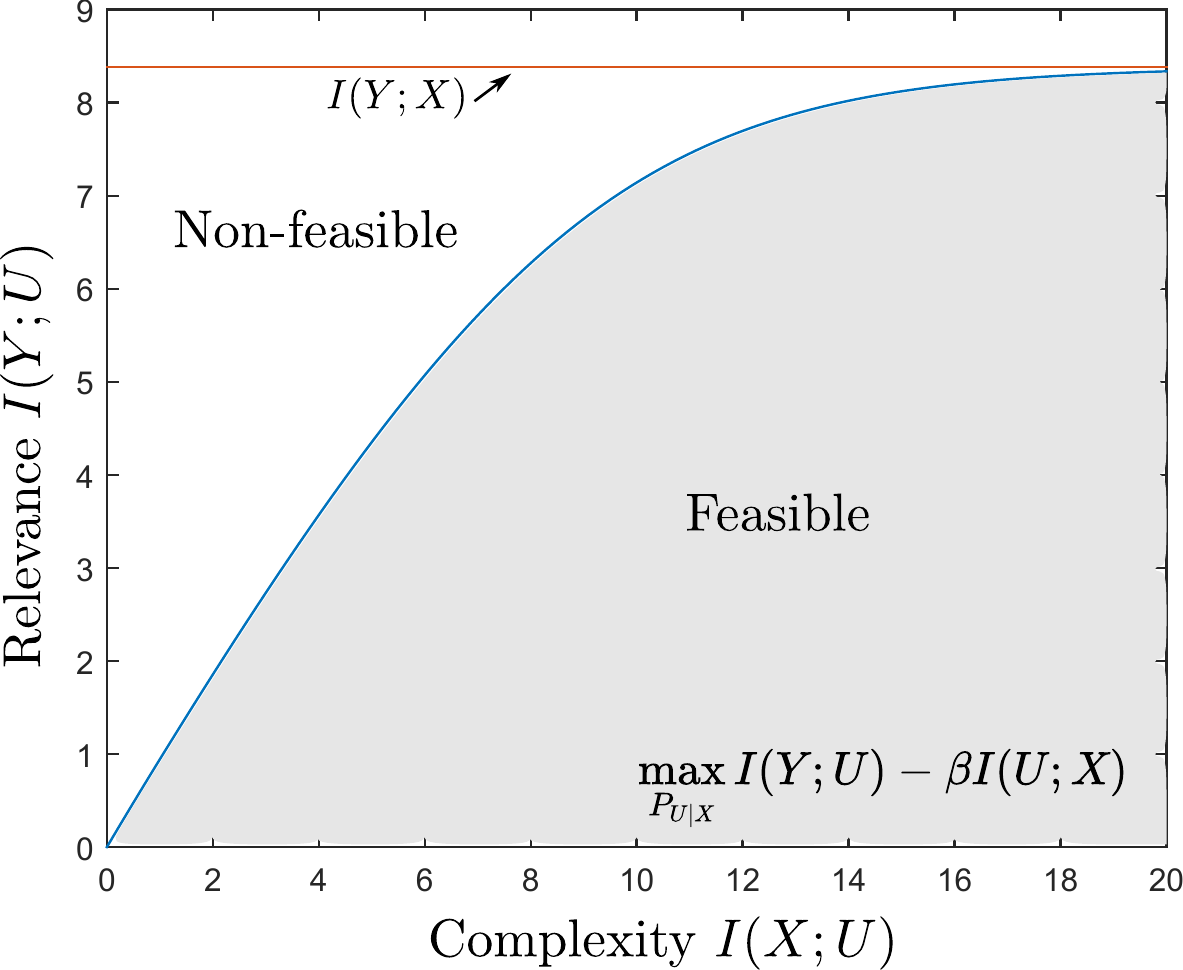}
\caption{Information bottleneck relevance--complexity region. For a given $\beta$, the solution $P^{*,\beta}_{U|X}$ to the minimization of the IB-Lagrangian in Equation         \eqref{IB-Lagrangian-formulation-minimizing-complexity} results in a pair $(\Delta_{\beta}, R_{\beta})$ on the boundary of the IB relevance--complexity region (in grey).} 
\label{fig-region}
\end{figure}

The minimization of the IB-Lagrangian $\mc L_{\beta}$ in Equation         \eqref{eq:IBCriteria} for a given $\beta\geq 0$ and $P_{X,Y}$ results in an optimal mapping $P^{*,\beta}_{U|X}$ and a relevance--complexity pair $(\Delta_{\beta}, R_{\beta})$ where $\Delta_{\beta}=I(U_{\beta}, X) $ and $R_{\beta}=I(U_{\beta}, Y) $ are, respectively, the relevance and the complexity resulting from generating $U_{\beta}$ with the solution $P^{*,\beta}_{U|X}$. 
By optimizing over all $\beta\geq 0$, the resulting relevance--complexity pairs $(\Delta_{\beta}, R_{\beta})$ characterize the boundary of the region of simultaneously achievable relevance--complexity pairs for a distribution  $P_{X,Y}$ (see Figure~\ref{fig-region}). In particular, for a fixed $P_{X,Y}$, we define this region as the union of relevance--complexity pairs $(\Delta, R)$ that satisfy
\begin{align}
\Delta\leq I(U,Y), \quad R\geq I(X,U)
\end{align}
where the union is over all $P_{U|X}$ such that   $U$ satisfies    $U \mkv X \mkv Y$ form a Markov Chain in this order. Any pair $(\Delta, R)$ outside of this region is not simultaneously achievable by any mapping $P_{U|X}$.

\section{Solutions to the Information Bottleneck Problem}\label{sec:SolutionsIB}

As shown in the previous region, the IB problem provides a methodology to design mappings $P_{U|X}$ performing at different relevance--complexity points within the region of feasible $(\Delta, R)$ pairs, characterized by the IB relevance--complexity region, 
by minimizing the IB-Lagrangian in Equation~\eqref{IB-Lagrangian-formulation-minimizing-complexity} for different values of $\beta$. However, in general, this optimization is challenging as it requires computation of mutual information terms.

 In this section, we describe how, for a fixed parameter $\beta$, the optimal solution $P^{\beta, *}_{U|X}$, or an efficient approximation of it, can be obtained under: (i) particular distributions, e.g., Gaussian and  binary symmetric sources; (ii) known general discrete memoryless distributions; and (iii) unknown memory distributions and only samples are available.

\subsection{Solution for Particular Distributions: Gaussian and Binary Symmetric Sources}
In certain cases, when the joint distribution $P_{X,Y}$ is know, e.g., it is binary symmetric or Gaussian, information theoretic inequalities can be used to minimize the IB-Lagrangian in \eqref{eq:IBCriteria} in closed form.  

\subsubsection{Binary IB}
 Let $X$ and $Y$ be a doubly symmetric binary sources (DSBS), i.e.,  $(X,Y) \sim \text{DSBS}(p)$ for some $0 \leq p \leq 1/2$.  ({A DSBS is a pair $(X,Y)$ of binary random variables $X\sim\text{Bern}(1/2)$ and $Y\sim\text{Bern}(1/2)$ and $X\oplus Y\sim\text{Bern}(p)$, where $\oplus$ is the sum modulo $2$. That is, $Y$ is the output of a binary symmetric channel with crossover probability $p$ corresponding to the input $X$, and $X$ is the output of the same channel with input $Y$.) Then, it can be shown that the optimal $U$ in  \eqref{eq:IBCriteria}  is such that $(X,U) \sim \text{DSBS}(q)$ for some $0 \leq q \leq 1$. Such a $U$ can be obtained with the mapping $P_{U|X}$ such that
\begin{align}
U=X\oplus Q, \quad\text{with } Q\sim \text{DSBS}(q).
\end{align}

 In this case, straightforward algebra leads to that the complexity level is given by
\begin{equation}
I(U;X) = 1 - h_2(q),
\end{equation}
where, for $0 \leq x \leq 1$, $h_2(x)$ is the entropy of a Bernoulli-$(x)$ source, i.e., $h_2(x) = - x \log_2(x) - (1-x)\log_2(1-x)$, and the relevance level is given by
\begin{equation}
I(U;Y) = 1-h_2(p \star q)
\end{equation}
where $p \star q = p(1-q) + q(1-p)$. The result extends easily to discrete symmetric mappings $Y \longrightarrow X$ with binary $X$ (one bit output quantization) and discrete non-binary  $Y$.

\subsubsection{Vector Gaussian IB}\label{sssec:GaussianIB}

 Let $(\dv X, \dv Y) \in \mathbb{C}^{N_x} \times \mathbb{C}^{N_y}$ be a pair of jointly Gaussian, zero-mean, complex-valued random vectors, of dimension $N_x> 0$ and $N_y>0$, respectively. In this case, the optimal solution of the IB-Lagrangian in Equation         \eqref{IB-Lagrangian-formulation-minimizing-complexity}  (i.e., test channel $P_{U|X}$) is a noisy linear projection to a subspace whose dimensionality is determined by the tradeoff parameter $\beta$. The subspaces are spanned by basis vectors in a manner similar to the well known canonical correlation analysis~\cite{H35}. For small $\beta$, only the vector associated to the dimension with more energy, i.e., corresponding to the largest eigenvalue of a particular hermitian matrix, will be considered in $U$. As $\beta$ increases, additional dimensions are added to $U$ through a series of critical points that are similar to structural phase transitions. This process continues until $U$ becomes rich enough to capture all the relevant information about $Y$ that is contained in $X$. 
In particular, the boundary of the optimal relevance--complexity region was shown in~\cite{GT04} to be achievable using a test channel $P_{\dv U|\dv X}$, which is such that $(\dv U,\dv X)$ is Gaussian. Without loss of generality, let 
\begin{equation}\label{eq:GaussianTest}
\dv U = \dv A \dv X + \boldsymbol{\xi} 
\end{equation}
where $\dv A \in \mc M_{N_u,N_x}(\mathbb{C})$ is an $N_u \times N_x$ complex valued matrix and $\boldsymbol{\xi} \in \mathbb{C}^{N_u}$ is a Gaussian noise that is independent of $(\dv X,\dv Y)$ with zero-mean and covariance matrix $\dv I_{N_u}$. For a given non-negative trade-off parameter $\beta$, the matrix $\dv A$ has a number of rows that depends on $\beta$ and is given by~\cite[Theorem 3.1]{CGTW05}
\begin{equation}
\dv A = \left\{
\begin{array}{cc}
\left[\mathbf{0}^T; \dots; \mathbf{0}^T\right], &  0 \leq \beta < \beta^{c}_1 \\
\left[ \alpha_{1} \dv v_1^T; \mathbf{0}^T; \dots; \mathbf{0}^T\right], & \beta^{c}_1 \leq \beta < \beta^{c}_2 \\
\left[ \alpha_{1} \dv v_1^T;\alpha_{2} \dv v_2^T; \mathbf{0}^T; \dots; \mathbf{0}^T\right], & \beta^{c}_2 \leq \beta < \beta^{c}_3 \\
\vdots & 
\end{array}
\right\}
\end{equation}
where $\{\dv v_1^T,  \dv v_2^T,  \hdots,  \dv v_{N_x}^T \}$ are the left eigenvectors of $\boldsymbol{\Sigma}_{\dv X|\dv Y}\boldsymbol{\Sigma}_{\dv X}^{-1}$ sorted by their corresponding ascending eigenvalues $\lambda_1, \lambda_2, \hdots, \lambda_{N_x}$. Furthermore, for $i=1,\hdots,N_x$, $\beta_i^c = \frac{1}{1-\lambda_i}$ are critical $\beta$-values, $\alpha_i = \sqrt{\frac{\beta(1-\lambda_i)-1}{\lambda_i r_i}}$ with $r_i =  \dv v_i^T \boldsymbol{\Sigma}_{\dv X} \dv v_i$, $\mathbf{0}^T$ denotes the $N_x$-dimensional zero vector and semicolons separate the rows of the matrix. 
\noindent It is interesting to observe that the optimal projection consists of eigenvectors of $\boldsymbol{\Sigma}_{\dv X|\dv Y}\boldsymbol{\Sigma}_{\dv X}^{-1}$, combined in a judicious manner: for values of $\beta$ that are smaller than $\beta^c_1$, reducing complexity is of prime importance, yielding extreme compression $\dv U=\boldsymbol{\xi}$, i.e., independent noise and no information preservation at all about $\dv Y$. As $\beta$ increases, it undergoes a series of critical points $\{\beta^c_i\}$, at each of which a new eignevector is added to the matrix $\dv A$, yielding a more complex but richer representation---the rank of $\dv A$ increases accordingly.

For the specific case of scalar Gaussian sources, that is $N_x=N_y=1$, e.g., $X = \sqrt{\text{snr}} Y + N$ where $N$ is standard Gaussian with zero-mean and unit variance, the above result simplifies considerably. In this case, let without loss of generality the mapping  $P_{U|X}$ be given by
\begin{equation}
X = \sqrt{a} X + Q
\end{equation} 
where $Q$ is standard Gaussian with zero-mean and  variance $\sigma_q^2$. In this case, for $I(U;X)=R$, we get
\begin{equation}
I(U;Y) = \frac{1}{2} \log(1+\text{snr}) - \frac{1}{2}\log\Big(1 + \text{snr} \exp(-2R)\Big).
\end{equation}

\subsection{Approximations for Generic Distributions}

Next, we present an approach to obtain solutions to the the information bottleneck problem for generic distributions,   both when    this solution is known and when it is unknown. The method consists in defining a variational (lower) bound on the IB-Lagrangian, which can be optimized more easily than optimizing the IB-Lagrangian directly.

\subsubsection{A Variational Bound}

Recall the IB goal of finding a representation $U$ of $X$ that is maximally informative about $Y$ while being concise enough (i.e., bounded $I(U;X)$). This corresponds to optimizing the IB-Lagrangian
\begin{equation}
\mc L_{\beta}^{\mathrm{IB}}(P_{U|X}):= I(U;Y) - \beta I(U;X)
\label{IB-Lagrangian-formulation-maximizing-relevance-2}
\end{equation}
where the maximization is over all stochastic mappings $P_{U|X}$ such that $U \mkv X \mkv Y$ and $|\mc U| \leq |\mc X|+1$. In this section, we   show that minimizing Equation         \eqref{IB-Lagrangian-formulation-maximizing-relevance-2} is equivalent to optimizing the variational cost
\begin{equation}
\mc L^{\mathrm{VIB}}_{\beta} (P_{U|X},Q_{Y|U},S_{U}):=\mathrm{E}_{P_{U|X}}\left[ \log Q_{Y|U}(Y|U)\right ]  - \beta D_{\mathrm{KL}}(P_{U|X}|S_{U}),
\label{variational-bound-IB-problem}
\end{equation}
where $Q_{Y|U}(y|u)$ is an given stochastic map $Q_{Y|U} \: :\: \mc U \rightarrow [0,1]$ (also referred to as the  variational approximation of $P_{Y|U}$ or decoder) and $S_{U}(u):\mc U\rightarrow [0,1]$ is a given stochastic map (also referred to as the  variational approximation of $P_U$), and  $D_{\mathrm{KL}}(P_{U|X}|S_{U})$ is the relative entropy between $P_{U|X}$ and $S_U$.

Then, we have the following bound for a any valid $P_{U|X}$, i.e., satisfying the Markov Chain    in~\eqref{MK-chain},
\begin{align}
\mc L_{\beta}^{\mathrm{IB}}(P_{U|X})\geq \mc L_{\beta}^{\mathrm{VIB}}(P_{U|X}, Q_{Y|U},S_{U} ),\label{eq:InequatlityVariational}
\end{align}
where the equality holds when $Q_{Y|U}= P_{Y|U}$ and $S_{U} = P_U$, i.e., the variational approximations correspond to the true value.

In the following,    we derive the variational bound. Fix $P_{U|X}$ (an encoder)  and the variational decoder approximation  $Q_{Y|U}$. The relevance $I(U;Y)$ can be lower-bounded as
\begin{align}
I(U;Y) &= \displaystyle\sumint_{u \in \mc U,\: y \in \mc Y} P_{U,Y}(u,y) \log \frac{P_{Y|U}(y|u)}{P_Y(y)} d_yd_u \\
& \stackrel{(a)}{=} \displaystyle\sumint_{u \in \mc U,\: y \in \mc Y} P_{U,Y}(u,y) \log \frac{Q_{Y|U}(y|u)}{P_Y(y)} d_yd_u
 + D\Big(P_Y\|Q_Y|U\Big) \\
& \stackrel{(b)}{\geq} \displaystyle\sumint_{u \in \mc U,\: y \in \mc Y} P_{U,Y}(u,y) \log \frac{Q_{Y|U}(y|u)}{P_Y(y)} d_yd_u \\
&= H(Y) + \displaystyle\sumint_{u \in \mc U,\: y \in \mc Y} P_{U,Y}(u,y) \log Q_{Y|U}(y|u) d_yd_u \\
& \stackrel{(c)}{\geq} \displaystyle\sumint_{u \in \mc U,\: y \in \mc Y} P_{U,Y}(u,y) \log Q_{Y|U}(y|u) d_yd_u \\
& \stackrel{(d)}{=} \displaystyle\sumint_{u \in \mc U,\: x \in \mc X,\: y \in \mc Y} P_X(x)P_{Y|X}(y|x)P_{U|X}(u|x) \log Q_{Y|U}(y|u) d_xd_yd_u,
\label{variational-bound-relevance-term}
\end{align} 
where in $(a)$ the term $D\Big(P_Y\|Q_Y|U\Big)$ is the conditional relative entropy between $P_Y$ and $Q_Y$, given $P_U$; $(b)$ holds by the non-negativity of relative entropy; $(c)$ holds by the non-negativity of entropy; and $(d)$ follows using the Markov Chain $U \mkv X \mkv Y$.

\noindent Similarly, let $S_U$  be a given the variational approximation of $P_U$. Then, we get
\begin{align}
I(U;X) &= \displaystyle\sumint_{u \in \mc U,\: x \in \mc X} P_{U,X}(u,x) \log \frac{P_{U|X}(u|x)}{P_U(u)} d_xd_u \\ 
&= \displaystyle\sumint_{u \in \mc U,\: x \in \mc X} P_{U,X}(u,x) \log \frac{P_{U|X}(u|x)}{S_U(u)} d_xd_u - D\Big(P_U\|S_U\Big) \\
& \leq \displaystyle\sumint_{u \in \mc U,\: x \in \mc X} P_{U,X}(u,x) \log \frac{P_{U|X}(u|x)}{S_U(u)} d_xd_u 
\label{variational-bound-complexity-term}
\end{align}
 where the inequality follows since the relative entropy is non-negative.
 
  Combining Equations         \eqref{variational-bound-relevance-term} and           \eqref{variational-bound-complexity-term}, we get
\begin{align}
I(U;Y) - \beta I(U;X) &\geq \displaystyle\sumint_{u \in \mc U,\: x \in \mc X,\: y \in \mc Y} P_X(x)P_{Y|X}(y|x)P_{U|X}(u|x) \log Q_{Y|U}(y|u) d_xd_yd_u \nonumber\\
& \quad - \beta \displaystyle\sumint_{u \in \mc U,\: x \in \mc X} P_{U,X}(u,x) \log \frac{P_{U|X}(u|x)}{S_U(u)} d_xd_u.
\end{align}

The use of the variational bound   in          \eqref{variational-bound-IB-problem} over the IB-Lagrangian   in         \eqref{IB-Lagrangian-formulation-maximizing-relevance-2} shows some advantages. First,  it allows the derivation of alternating algorithms that allow to obtain a solution by optimizing over the encoders and decoders. Then, it is easier to obtain an empirical estimate of Equation         \eqref{variational-bound-IB-problem} by sampling from: (i) the joint distribution $P_{X,Y}$; (ii) the encoder $P_{U|X}$; and (iii) the prior $S_U$.  
Additionally, as noted in          \eqref{eq:InequatlityVariational}, when evaluated for the optimal decoder $Q_{Y|U}$ and prior $S_U$, the variational bound becomes tight. All this allows   obtaining  algorithms to obtain good approximate solutions to the IB problem, as shown next.
Further theoretical implications of this variational bound are discussed in \cite{aleks2019difference}.

\subsubsection{Known Distributions}
Using the variational formulation in       \eqref{variational-bound-IB-problem}, when the data model is discrete and the joint distribution $P_{X,Y}$ is known,  the IB problem can be solved by using an
 iterative method that optimizes the variational IB cost function in Equation         \eqref{variational-bound-IB-problem}
alternating over the distributions $P_{U|X}, Q_{Y|U}$, and $S_{U}$.
In this case, the maximizing distributions $P_{U|X}, Q_{Y|U}$, and $S_{U}$ can be
efficiently found by an alternating optimization procedure similar to the expectation-maximization
(EM) algorithm \cite{Dempster1977} and the standard Blahut--Arimoto (BA) method \cite{B72}.
In particular, a solution $P_{U|X}$ to the constrained optimization problem is determined by the following self-consistent equations, for all $(u,x,y) \in \mc U \times \mc X \times \mc Y$, ~\cite{TPB99} 
\begin{subequations}
\begin{align}
P_{U|X}(u|x) &= \frac{P_U(u)}{Z(\beta,x)} \exp \Big(-\beta D_{\mathrm{KL}}\Big(P_{Y|X}(\cdot|x)\|P_{Y|U}(\cdot|u)\Big)\Big) \\
P_U(u) &= \sum_{x \in \mc X} P_X(x)P_{U|X}(u|x) \\
P_{Y|U}(y|u) &= \sum_{x \in \mc X} P_{Y|X}(y|x)P_{X|U}(x|u)
\end{align}
\label{self-consistent-equations-IB-algorithm}
\end{subequations} 
where $P_{X|U}(x|u) = P_{U|X}(u|x)P_X(x)/P_U(u)$ and $Z(\beta,x)$ is a normalization term. It is shown in~\cite{TPB99} that alternating iterations of these equations converges to a solution of the problem for any initial $P_{U|X}$. However, by opposition to the standard {Blahut--Arimoto} algorithm~\cite{B72,A72}, which is classically used in the computation of rate-distortion functions of discrete memoryless sources for which convergence to the optimal solution is guaranteed, convergence here may be to a local optimum only. If $\beta = 0$. the optimization is non-constrained and one can set $U=\emptyset$, which yields minimal relevance and complexity levels. Increasing the value of $\beta$ steers towards more accurate and more complex representations, until $U=X$ in the limit of very large (infinite) values of $\beta$ for which the relevance reaches its maximal \mbox{value $I(X;Y)$. } 
For discrete sources with (small) alphabets, the updating   equations described by Equation         \eqref{self-consistent-equations-IB-algorithm}  are relatively
easy computationally. However,  if the variables $X$ and $Y$ lie in a continuum, solving the equations described by Equation         \eqref{self-consistent-equations-IB-algorithm} is very challenging. In the case in which $X$ and $Y$ are joint multivariate Gaussian, the problem of finding the optimal representation $U$ is analytically tractable in~\cite{GT04} (see also the related~\cite{CGTW05,WM14}), as discussed in Section~\ref{sssec:GaussianIB}.
 Leveraging   the optimality of Gaussian mappings $P_{U|X}$   to restrict the optimization of $P_{U|X}$ to
Gaussian distributions as     in Equation         \eqref{eq:GaussianTest},   allows   reducing the search of update rules to those of the associated parameters,
namely covariance matrices. When {$Y$ is a deterministic function of $X$}, the IB curve cannot be explored, and other Lagrangians have been proposed to tackle this problem~\cite{glvez2019convex}. 
\subsection{Unknown Distributions}

The main drawback of the solutions presented thus far for the IB principle is that, in the exception of small-sized discrete $(X,Y)$ for which iterating Equation         \eqref{self-consistent-equations-IB-algorithm} converges to an (at least local) solution and jointly Gaussian $(X,Y)$ for which an explicit analytic solution was found, solving Equation         \eqref{IB-Lagrangian-formulation-minimizing-complexity} is generally computationally costly, especially for high dimensionality.  Another important barrier in solving Equation         \eqref{IB-Lagrangian-formulation-minimizing-complexity} directly is that IB necessitates knowledge of the joint distribution $P_{X,Y}$. In this section, we describe a method to provide an approximate solution to the IB problem in the case in which the joint distribution is unknown and only a give training set of $N$ samples $\{(x_i,y_i)\}_{i=1}^N$ is available.

 A major step ahead, which widened   the range of applications of IB inference for various learning problems, appeared in~\cite{AFDM17}, where the authors used neural networks to parameterize the variational inference lower bound   in  Equation         \eqref{variational-bound-IB-problem} and show that its optimization can be done through the classic and widely used stochastic gradient descendent (SGD).  This method, denoted by Variational IB in~\cite{AFDM17} and detailed below, has allowed   handling  handle high-dimensional, possibly continuous, data, even in the case in which the distributions are unknown.  

\subsubsection{Variational IB}\label{sssec:VIB}
The goal of the variational IB  when only   samples $\{(x_i,y_i)\}_{i=1}^N$ are available is to solve the IB problem optimizing an approximation of the cost function. 
 For instance, for a given training set $\{(x_i,y_i)\}_{i=1}^N$, the right hand side of Equation \eqref{variational-bound-IB-problem} can be approximated as
\begin{equation}
\mc L_{\text{low}} \approx \frac{1}{N}\sum_{i=1}^N \left[ \displaystyle\sumint_{u \in \mc U} \Big(P_{U|X}(u|x_i)\log Q_{Y|U}(y_i|u) - \beta P_{U|X}(u|x_i) \log\frac{P_{U|X}(u|x_i)}{S_U(u)} \Big) d_u \right].
\end{equation}

However, in general, the direct optimization of this cost is challenging. In the variational IB method, this optimization is done  by parameterizing the encoding and decoding distributions $P_{U|X}$, $Q_{Y|U} $, and $S_U$ that are to optimize using a family of distributions whose parameters are determined by DNNs. This
allows us to formulate  Equation         \eqref{variational-bound-IB-problem}  in terms of the DNN parameters, i.e., its weights, and optimize it by using the reparameterization
trick~\cite{KW13}, Monte Carlo sampling, and  stochastic gradient descent (SGD)-type algorithms.

Let $P_{\theta}(u|x)$ denote the family of encoding probability distributions $P_{U|X}$ over $\mc U$ for each element on $\mc X$, parameterized by the output of a DNN $f_{\theta}$ with parameters $\theta$.
A common example  is the family of multivariate Gaussian distributions \cite{KW13}, which are parameterized by the mean $\boldsymbol{\mu}^{\theta}$ and covariance matrix $\dv \Sigma^{\theta}$, i.e., $\boldsymbol\gamma:= (\boldsymbol{\mu}^{\theta}, \dv \Sigma^{\theta})$.  Given an observation $X$, the values of $(\boldsymbol{\mu}^{\theta}(x), \dv \Sigma^{\theta}(x))$ are determined by the output of the DNN $f_{\theta}$, whose input is $X$, and the corresponding family member is given by $P_{\theta}(u|x) = \mc {N} (u; \boldsymbol{\mu}^{\theta}(x), \dv \Sigma^{\theta}(x))$.  For discrete distributions, a common example are concrete variables \cite{Jang2017} (or Gumbel-Softmax \cite{Maddison2016}). Some details are given below.

Similarly, for decoder $Q_{Y|U}$ over $\mc Y$ for each element on $\mc U$, let $Q_{\psi}(y|u)$ denote the family of distributions parameterized by the output of the DNNs $f_{\psi_k}$. Finally, for the prior distributions $S_{U}(u)$ over $\mc U$ we define the family of distributions $S_{\varphi}(u)$, which do not depend on a DNN.

By restricting the optimization of the variational IB cost in Equation         \eqref{variational-bound-IB-problem} to the encoder, decoder,  and prior within the families of distributions  $P_{\theta}(u|x)$, $Q_{\psi}(y|u)$, and $S_{\varphi}(u)$, we get
\begin{align}
\max_{ P_{U|X}}\max_{ Q_{Y|U},S_U}\mc L^{\mathrm{VIB}}_{\beta} (P_{U|X},Q_{Y|U},S_{U})\geq \max_{\theta,\phi, \varphi}\mc L^{\mathrm{NN}}_{\beta}({\theta, \phi,  \varphi} ),\label{eq:NNCostBound}
\end{align}
where  $\theta,\phi$, and $\varphi$   denote the DNN parameters,   e.g., its weights, and  the cost in Equation         \eqref{eq:NNCost} is given~by
\begin{align}\label{eq:NNCost}
&\mc L^{\mathrm{NN}}_{\beta}({\theta, \phi,  \varphi} ):=
\mathds{E}_{P_{Y,X}}\mathds{E}_{\{P_{\theta}(U|X)\}}
\Big[\log Q_{\phi(Y| U)}(Y|U)\Big] -\beta D_{\mathrm{KL}}(P_{\theta}(U|X)\|S_{\varphi}(U)).
\end{align}


Next, using the training samples $\{(x_i,y_i)\}_{i=1}^N$, the DNNs are trained to maximize a Monte Carlo approximation of Equation         \eqref{eq:NNCost} over ${\boldsymbol\theta, \boldsymbol\phi, \boldsymbol \varphi}$ using optimization methods such as SGD or ADAM~\cite{Kingma2014} with backpropagation. 
However, in general, the direct computation of the gradients of Equation         \eqref{eq:NNCost} is challenging due to the dependency of the averaging with respect    to the encoding $P_{\theta}$, which makes it hard to approximate the cost by sampling. To circumvent this problem, the reparameterization trick~\cite{KW13}  is used to sample from $P_{\theta}(U|X)$. In particular, consider $P_{\theta}(U|X)$ to belong to a parametric family of distributions that can be sampled by first sampling a random variable $Z$ with distribution $P_{Z}(z)$, $z\in \mc Z$ and then transforming the samples using some function $g_{\theta}:\mc X\times \mc Z\rightarrow \mc U$ parameterized by $\theta$, such that $U = g_{\theta}(x,Z)\sim P_{\theta}(U|x)$.  Various parametric families of distributions fall within this class for both   discrete and continuous latent spaces, e.g., the Gumbel-Softmax distributions and the Gaussian distributions. Next, we detail how to sample from both examples:
\begin{enumerate}
\item Sampling from Gaussian Latent Spaces: When the latent space is a continuous vector space of dimension $D$, e.g., $\mc U=\mathds{R}^{D}$, we can
consider multivariate Gaussian parametric encoders of mean  $(\boldsymbol{\mu}^{\theta}$, and covariance $\dv \Sigma^{\theta})$, i.e.,  $P_{\theta}(u|x) = \mc {N} (u; \boldsymbol{\mu}^{\theta}, \dv \Sigma^{\theta})$. To sample $\dv U\sim \mc {N} (u; \boldsymbol{\mu}^{\theta}, \dv \Sigma^{\theta})$,  where 
 $\boldsymbol{\mu}^{\theta}(x)=f_{e,\theta}^{\mu}(x)$ and  $\dv \Sigma^{\theta}(x) = f^{\Sigma}_{e,\theta}(x)$  are determined as the output of a NN, sample a random variable $Z \sim \mc {N} (z; \dv 0, \dv I)$ i.i.d. and, given data sample $x\in \mc X$,  and generate the $j$th sample as
\begin{equation}
u_j = f_{e,\theta}^{\mu}(x) + f^{\Sigma}_{e,\theta}(x)  z_j
\end{equation}
where $z_j$ is a sample of $ Z\sim \mc N(\dv 0,\dv I)$ {, which is an independent Gaussian noise}, and $f^{\mu}_e(x)$ and $f^{\Sigma}_e(x)$ are the output values of the NN with weights $\theta$ for the given input sample $x$. 

An example of the resulting DIB architecture to optimize with an encoder, a latent space, and a decoder  parameterized by Gaussian distributions is shown in Figure~\ref{fig:enc_dec}. 

\item Sampling from a discrete latent space with the Gumbel-Softmax: 

If $U$ is categorical random variable on the finite set $\mathcal{U}$ of size $D$ with
probabilities $\pi := (\pi_1,\dots,   \pi_D)$), we can encode it as $D$-dimensional one-hot vectors lying on the corners of the
$(D-1)$-dimensional simplex, $\Delta_{D-1}$. In general, costs functions involving sampling from categorical distributions
are non-differentiable. Instead, we consider Concrete variables \cite{Maddison2016} (or Gumbel-Softmax \cite{Jang2017}), which are continuous
differentiable relaxations of categorical variables on the interior of the simplex, and are easy to sample. To sample
from a Concrete random variable $U\in \Delta_{D-1}$ 
at temperature $\lambda\in (0,\infty)$,   with probabilities $\pi\in (0,1)^{D}$, sample
$G_d\sim \mathrm{Gumbel}(0,1)$  i.i.d. {(}The $\mathrm{Gumbel}(0,1)$ distribution can be sampled by drawing $u\sim \mathrm{Uniform}(0,1)$ and calculating $g = -\log(-\log(u)) $. {)}, and set for each of the components of $U=(U_1,\ldots,U_D)$ 
\begin{align}
U_d = \frac{\exp(( \log(\pi_d+ G_d)/\lambda ))}{\sum_{j=1}^D\exp(( \log(\pi_j+ G_j)/\lambda ))},\quad d=1,\ldots, D.
\end{align} 
We denote by $Q_{\pi,\lambda(u,x)}$ the Concrete distribution with parameters $ (\pi(x), \lambda)$. When the temperature $\lambda$
approaches $0$, the samples from the concrete distribution become one-hot and $\mathrm{Pr}\{ \lim_{\lambda\rightarrow 0}U_{d}\}=\pi_d$ \cite{Jang2017}. Note that, for discrete data models, the application of Caratheodory's   theorem \cite{H-LKGC11} shows that the latent
variables $U$ that appear in  \eqref{IB-Lagrangian-formulation-minimizing-complexity} can be restricted     to   be with bounded alphabets size.
\end{enumerate}

\begin{figure}[!t]
	\centering
	\includegraphics[width=0.9\textwidth]{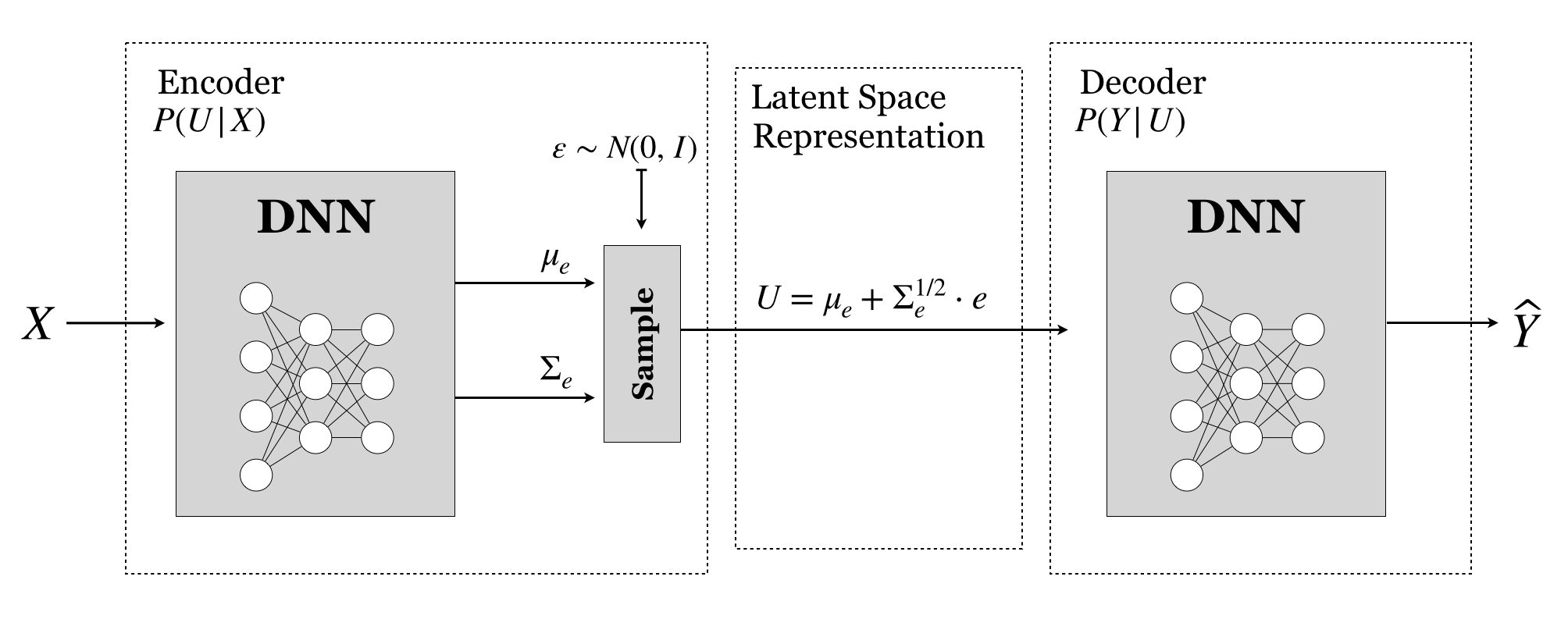}
	\caption{Example parametrization of {Variational Information B}ottleneck  using neural networks.} 
	\label{fig:enc_dec} 
\end{figure}

The reparametrization trick transforms the cost function  in  Equation         \eqref{eq:NNCost} into one which can be to approximated by sampling $M$ independent samples $\{u_{m}\}_{m=1}^M \sim P_{\theta}(u|x_{i})$ for each training sample $(x_{i},y_i)$, $i=1, \ldots, N$
 and allows   computing  estimates of the gradient using backpropagation~\cite{KW13}. Sampling is performed by using $u_{i,m} = g_{\phi}(x_{i},z_{m})$ with $\{z_{m}\}_{m=1}^M$ i.i.d. sampled from $P_{Z}$. Altogether,  we have the empirical-DIB cost for the $i$th sample in the training dataset:
\begin{align}
&\mathcal{L}^{\mathrm{emp}}_{\beta,i}({\theta, \phi, \varphi} )
:=\frac{1}{M}\sum_{m=1}^M\Big[\log Q_{\phi}(y_i| u_{i,m})\label{eq:variational objective2}
-\beta D_{\mathrm{KL}}(P_{\theta}(U_{i}|x_{i})\|Q_{\varphi}(U_{i}))
\Big)\Big]. 
\end{align}
Note that, for many distributions, e.g., multivariate Gaussian, the divergence $D_{\mathrm{KL}}(P_{\theta}(U_{i}|x_{i})\|Q_{\varphi}(U_{i}))$ can be evaluated in closed form. Alternatively, an empirical approximation can be considered. 

Finally, we maximize the empirical-IB cost over the DNN parameters  ${\theta},\phi,\varphi$  as,
\begin{align}
 \max_{\boldsymbol{\theta},\boldsymbol{\phi}, \boldsymbol\varphi}\frac{1}{N}\sum_{i=1}^N\mathcal{L}^{\mathrm{emp}}_{\beta,i}({\boldsymbol\theta, \boldsymbol\phi, \boldsymbol \varphi} ).\label{eq:Lhatoptimization}
\end{align}

By the law of large numbers, for large $N,M$, we have $1/N\sum_{i=1}^M\mathcal{L}^{\mathrm{emp}}_{\beta,i}({\theta, \phi,\varphi}) \rightarrow \mc L^{\mathrm{NN}}_{\beta}({\theta, \phi,\varphi} )$ almost surely.
After convergence of the DNN parameters  to ${\theta}^*,{\phi}^*, \varphi^*$,  for a new observation $X$, the representation $U$  can be obtained  by sampling from the encoders $P_{\theta_k^*}(U_k|X_k)$. In addition, note that a soft estimate of the remote source $Y$ can be inferred by sampling from the decoder $Q_{\phi^*}(Y|U)$. The notion of encoder and decoder in the IB-problem will come clear from its relationship with lossy source coding in Section~\ref{ssec:LossySourceCoding}.

\section{Connections to Coding Problems}

The IB problem is a one-shot coding problem, in the sense that the  operations are performed letter-wise. In this section, we consider now the relationship between the IB problem and  (asymptotic)  coding problem in which the coding operations are performed over blocks of size $n$, with $n$ assumed to be large and the joint distribution of the data $P_{X,Y}$ is in general assumed to be known a priori. 
The connections between these problems allow   extending  results from one setup to another, and to consider generalizations of the classical IB problem to other setups, e.g., as shown in Section~\ref{sec:Distributed-Information-Bottleneck}.

\subsection{Indirect Source Coding under Logarithmic Loss}\label{ssec:LossySourceCoding}

Let us consider the (asymptotic) indirect source coding problem shown in Figure~\ref{fig-remote-source-coding-problem}, in which $Y$ designates a memoryless remote source and $X$ a noisy version of it that is observed at the encoder. 

\begin{figure}[H]
\centering
\includegraphics[width=0.95\textwidth]{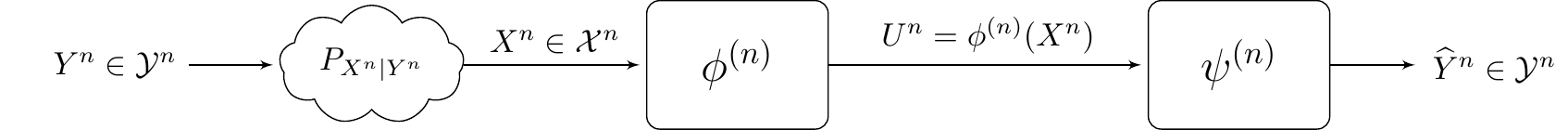}
\caption{A remote source coding problem.} 
\label{fig-remote-source-coding-problem}
\end{figure}
A sequence of $n$ samples $X^n=(X_1,\ldots, X_n)$ is mapped by an encoder $\phi^{(n)}:\mc X^n\rightarrow \{1,\ldots, 2^{nR}\}$ which outputs a message from a set $\{1, \ldots, 2^{nR}\}$, that is, the encoder uses at most $R$ bits per sample to describe its observation and the range of the encoder map is allowed to grow with the size of the input sequence as   
\begin{equation}
\|\phi^{(n)}\| \leq nR.
\label{remote-source-coding-input-constraint}
\end{equation}

This message is mapped with a decoder $\phi^{(n)}: \{1,\ldots, 2^{nR}\}\rightarrow \hat{\mathcal{Y}}$ to generate a reconstruction of the source sequence $Y^n$ as $\mathcal{Y}^n\in \hat{\mathcal{Y}}^n$.
 As already observed in~\cite{HT07}, the IB problem in Equation         \eqref{IB-Lagrangian-formulation-minimizing-complexity} is essentially equivalent to a remote point-to-point source coding problem in which distortion between $Y^n$ as $\mathcal{Y}^n\in \hat{\mathcal{Y}}^n$ is measured under the logarithm loss (log-loss) fidelity criterion \cite{CW14}. That is, rather than just assigning a deterministic value to each sample of the source, the decoder gives an assessment of
the degree of confidence or reliability on each estimate. Specifically, given the output description $m=\phi^{(n)}(x^n)$ of the encoder, the decoder generates a soft-estimate $\hat{y}^n$ of $y^n$ in the form of a probability distribution over $\mathcal{Y}^n$, i.e.,
$\hat{y}^n=\hat{P}_{Y^n|M}(\cdot)$. The incurred discrepancy between $y^n$ and the estimation $\hat{y}^n$ under log-loss for the observation $x^n$
is then given by the
per-letter logarithmic loss distortion, which is defined as
 \begin{equation}
\ell_{\text{log}}(y,\hat{y}) := \log \frac{1}{\hat{y}(y)}.
\label{definition-per-letter-log-loss-distortion-measure}
\end{equation}
for $y \in \mc Y$ and $\hat{y} \in \mc P(\mc Y)$ designates here a probability distribution on $\mc Y$ and $\hat{y}(y)$ is the value of that distribution evaluated at the outcome $y \in \mc Y$.

That is, the encoder uses at most $R$ bits per sample to describe its observation to a decoder which is interested in reconstructing the remote source $Y^n$ to within an average distortion level $D$, using a per-letter distortion metric,  i.e.,
\begin{equation}
\mathbb{E}[\ell_{\text{log}}^{(n)}(Y^n, \hat{Y}^n)] \leq D
\end{equation}
 where the incurred distortion between two sequences $Y^n$ and $\hat{Y}^n$ is measured as 
\begin{equation}
\ell_{\text{log}}^{(n)}(Y^n, \hat{Y}^n) = \frac{1}{n} \sum_{i=1}^{n} \ell_{\text{log}}(y_i,\hat{y}_i)
\label{definition-log-loss-distortion-measure-n-letter}
\end{equation}
 and the  per-letter distortion is measured in terms of  that  given by the logarithmic loss in Equation         \eqref{definition-per-letter-log-loss-distortion-measure}.

\noindent The rate distortion region of this model is given by the union of all pairs $(R,D)$ that satisfy~\cite{DT62,W80}
\begin{subequations}
\begin{align}
 R &\geq I(U;X) \\
 D &\geq H(Y|U)
\end{align} 
\end{subequations}
where the union is over all auxiliary random variables $U$ that satisfy that $ U \mkv X \mkv Y$ forms a Markov Chain in this order. Invoking the support lemma~\cite[p. 310]{CK81}, it is easy to see that this region is not altered if one restricts $U$ to satisfy $|\mc U| \leq |\mc X|+1$. In addition, using the substitution $\Delta := H(Y) - D$, the region can be written equivalently as the union of all pairs $(R,H(Y)-\Delta)$ that satisfy
\begin{subequations}
\begin{align}
 R &\geq I(U;X) \\
 \Delta &\leq I(U;Y)
\end{align} 
\end{subequations}
where the union is over all $U$s with pmf $P_{U|X}$ that satisfy $U \mkv X \mkv Y$, with $|\mc U| \leq |\mc X|+1$.

The boundary of this region is equivalent to the one described by the IB principle in Equation         \eqref{IB-Lagrangian-formulation-minimizing-complexity} if solved for all $\beta$, and therefore the IB problem is essentially a remote source coding problem in which the distortion is measured under the logarithmic loss measure. Note that, operationally, the IB problem is equivalent to that of finding an encoder $P_{U|X}$ which maps the observation $X$ to a representation $U$ that satisfies the bit rate constraint $R$ and such that $U$ captures enough relevance of $Y$ so that the posterior probability of $Y$ given $U$ satisfies an average distortion constraint.

\subsection{Common Reconstruction}

Consider the problem of source coding with side information at the decoder, i.e., the well known Wyner--Ziv setting~\cite{WZ76}, with the distortion measured under logarithmic-loss. Specifically, a memoryless source $X$ is to be conveyed lossily to a decoder that observes a statistically correlated side information $Y$. The encoder uses $R$ bits per sample to describe its observation to the decoder which wants to reconstruct an estimate of $X$ to within an average distortion level $D$, where the distortion is evaluated under the log-loss distortion measure. The rate distortion region of this problem is given by the set of all pairs $(R,D)$ that satisfy
\begin{equation} 
R + D \geq H(X|Y).
\end{equation}

The optimal coding scheme utilizes standard Wyner--Ziv compression~\cite{WZ76} at the encoder and the decoder map $\psi \: : \: \mc U \times \mc Y \to \hat{\mc X}$ is given by
\begin{equation}
\psi(U,Y) = \text{Pr}[X=x|U,Y]
\label{optimal-reproduction-function-WZ-setup-under-log-loss}
\end{equation}
for which it is easy to see that
\begin{equation}
\mathbb{E}[\ell_{\text{log}}(X,\psi(U,Y))] = H(X|U,Y).
\end{equation}

Now, assume that we constrain the coding in a manner that the encoder is be able to produce an exact copy of the compressed source constructed by the decoder. This requirement, termed \textit{common reconstruction} constraint (CR), was introduced and studied by Steinberg  ~\cite{S09} for various source coding models, including the Wyner--Ziv setup, in the context of a ``general distortion measure''. For the Wyner--Ziv problem under log-loss measure that is considered in this section, such a CR constraint causes some rate loss because the reproduction rule  in   Equation   \eqref{optimal-reproduction-function-WZ-setup-under-log-loss} is no longer possible. In fact, it is not difficult to see that under the CR constraint the above region reduces to the set of pairs 
 $(R,D)$ that satisfy
\begin{subequations}
\begin{align} 
\label{rate-distortion-region-WZ-setup-log-loss-and-CR-constraint-ineq1}
R &\leq I(U;X|Y) \\
D &\geq H(X|U)
\label{rate-distortion-region-WZ-setup-log-loss-and-CR-constraint-ineq2}
\end{align}
\end{subequations}
for some auxiliary random variable for which $U \mkv X \mkv Y$ holds. Observe that Equation         \eqref{rate-distortion-region-WZ-setup-log-loss-and-CR-constraint-ineq2} is equivalent to $I(U;X) \geq H(X) -D$ and that, for a given prescribed fidelity level $D$, the minimum rate is obtained for a description $U$ that achieves the inequality  in   Equation         \eqref{rate-distortion-region-WZ-setup-log-loss-and-CR-constraint-ineq2} with equality, i.e.,
\begin{equation}
R(D) = \min_{P_{U|X}\: : \: I(U;X)=H(X)-D} I(U;X|Y).
\end{equation}

\noindent Because $U \mkv X \mkv Y$, we have
\begin{equation}
I(U;Y) = I(U;X) - I(U;X|Y).
\end{equation}
Under the constraint $I(U;X)=H(X)-D$, it is easy to see that minimizing $I(U;X|Y)$ amounts to maximizing $I(U;Y)$, an aspect which bridges the problem at hand with the IB problem. 
  
 In the above, the side information $Y$ is used for binning but not for the estimation at the decoder. If the encoder ignores whether $Y$ is present or not at the decoder side, the benefit of binning is reduced---see the Heegard--Berger model with common reconstruction studied in~\cite{BZ16a,BZ16b}.


\subsection{Information Combining}
Consider again the IB problem. Assume one wishes to find the representation $U$ that maximizes the relevance $I(U;Y)$ for a given prescribed complexity level, e.g., $I(U;X)=R$. For this setup, we have
\begin{align}
I(X;U,Y) &= I(U;X) + I(Y;X) - I(U;Y) \\
        &= R  + I(Y;X) - I(U;Y)               
\end{align}
where the first equality holds since $U \mkv X \mkv Y$ is a Markov Chain. Maximizing $I(U;Y)$ is then equivalent to minimizing $I(X;U,Y)$. This is reminiscent of the problem of \textit{information combining}~\cite{SSZ05,LH06}, where $X$ can be interpreted as a source information that is conveyed through two channels: the channel $P_{Y|X}$ and the channel $P_{U|X}$. The outputs of these two channels are conditionally independent given $X$, and they should be processed in a manner such that, when combined, they preserve as much information as possible about $X$.

\subsection{Wyner--Ahlswede--Korner Problem}

Here, the two memoryless sources $X$ and $Y$ are encoded separately at rates $R_X$ and $R_Y$, respectively. A decoder gets the two compressed streams and aims at recovering $Y$ {losslessly}. This problem was studied and solved separately by Wyner~\cite{W75b} and Ahlswede and K\"orner~\cite{AK75}. For given $R_X=R$, the minimum rate $R_Y$ that is needed to recover $Y$ losslessly is
\begin{equation}
R^{\star}_Y(R) = \min_{P_{U|X}\: :\: I(U;X) \: \leq \: R} H(Y|U).
\end{equation}
Thus, we get
\begin{equation*}
\max_{P_{U|X}\: : \: I(U;X) \leq R} I(U;Y) = H(Y) - R^{\star}_Y(R),
\end{equation*}
and therefore, solving the IB problem is equivalent to solving the Wyner--Ahlswede--Korner Problem.
\subsection{The Privacy Funnel} 

Consider again the setting of Figure~\ref{fig-remote-source-coding-problem}, and let us assume that the pair $(Y,X)$ models data that a user possesses and which have the following properties: the data $Y$ are some sensitive (private) data that are not meant to be revealed at all, or else not beyond some level $\Delta$; and the data $X$ are non-private and are meant to be shared with another user (analyst). Because $X$ and $Y$ are correlated, sharing the non-private data $X$ with the analyst possibly reveals information about $Y$. For this reason, there is a trade off between the amount of information that the user shares about $X$ and the information that he keeps private about $Y$. The data $X$ are passed through a randomized mapping $\phi$ whose purpose is to make $U=\phi(X)$ maximally informative about $X$ while being minimally informative about $Y$. 

The analyst performs an inference attack on the private data $Y$ based on the disclosed information $U$. Let $\ell \: : \: \mc Y \times \hat{\mc Y} \longrightarrow \bar{\mathbb{R}}$ be an arbitrary loss function with reconstruction alphabet $\hat{\mc Y}$ that measures the cost of inferring $Y$ after observing $U$. Given $(X,Y) \sim P_{X,Y}$ and under the given loss function $\ell$, it is natural to quantify the difference between the prediction losses in predicting $Y \in \mc Y$ prior and after observing $U=\phi(X)$. Let
\begin{equation}
C(\ell,P) = \inf_{\hat{y} \in \hat{\mc Y}} \mathbb{E}_P[\ell(Y,\hat{y})] - \inf_{\hat{Y}(\phi(X))} \mathbb{E}_P[\ell(Y,\hat{Y})]
\label{inference-cost-gain-privacy-funnel}
\end{equation}
where $\hat{y} \in \hat{\mc Y}$ is deterministic and $\hat{Y}(\phi(X))$ is any measurable function of $U=\phi(X)$. The quantity $C(\ell,P)$ quantifies the reduction in the prediction loss under the loss function $\ell$ that is due to observing $U=\phi(X)$, i.e., the inference cost gain.  In~\cite{MSFM14} (see also ~\cite{ADAL19}), it is shown that that under some mild conditions the inference cost gain $C(\ell,P)$ as defined by Equation         \eqref{inference-cost-gain-privacy-funnel} is upper-bounded as
\begin{equation} 
C(\ell,P) \leq 2\sqrt{2} L \sqrt{I(U;Y)}
\label{upper-bound-on-inference-cost-gain-privacy-funnel}
\end{equation}
where $L$ is a constant. The inequality  in   Equation         \eqref{upper-bound-on-inference-cost-gain-privacy-funnel} holds irrespective to the choice of the loss function $\ell$, and this justifies the usage of the logarithmic loss function as given by Equation         \eqref{definition-per-letter-log-loss-distortion-measure} in the context of finding a suitable trade off between utility and privacy, since
\begin{equation}
I(U;Y) = H(Y) - \inf_{\hat{Y}(U)} \mathbb{E}_P[\ell_{\text{log}}(Y,\hat{Y})].
\end{equation}

 Under the logarithmic loss function, the design of the mapping $U=\phi(X)$ should strike a right balance between the utility for inferring the non-private data $X$ as measured by the mutual information $I(U;X)$ and the privacy metric about the private date $Y$ as measured by the mutual information $I(U;Y)$.  

\subsection{Efficiency of Investment Information}

Let $Y$ model a stock market data and $X$ some correlated information. In~\cite{EC98}, Erkip and Cover investigated how the description of the correlated information $X$ improves the investment in the stock market $Y$. Specifically, let $\Delta (C)$ denote the maximum increase in growth rate when $X$ is described to the investor at rate $C$. Erkip and Cover found a single-letter characterization of the incremental growth rate $\Delta(C)$. When specialized to the horse race market, this problem is related to the aforementioned source coding with side information of Wyner~\cite{W75b} and Ahlswede-K\"orner~\cite{AK75}, and, thus, also to the IB problem. The work~in \cite{EC98} provides explicit analytic solutions for two horse race examples, jointly binary and jointly Gaussian horse races.

\section{Connections to Inference and Representation Learning}
In this section, we consider the connections of the IB problem with learning, inference and generalization, for which, typically, the joint distribution $P_{X,Y}$ of the data is not known and only a set of samples is available.

\subsection{Inference Model} 

Let a measurable variable $X \in \mc X$ and a target variable $Y \in \mc Y$ with unknown joint distribution $P_{X,Y}$ be given. In the classic problem of statistical learning, one wishes to infer an accurate predictor of the target variable $Y \in \mc Y$ based on observed realizations of $X \in \mc X$. That is, for a given class $\mc F$ of admissible predictors $\psi\: : \: \mc X \to \hat{\mc Y}$ and a loss function $\ell : \mc Y \to \hat{\mc Y}$ that measures discrepancies between true values and their estimation, one aims at finding the mapping $\psi \in \mc F$ that minimizes the expected (population) risk 
\begin{equation}
\mc C_{P_{X,Y}}(\psi,\ell) = \mathbb{E}_{P_{X,Y}} [\ell(Y,\psi(X))].
\label{definition-population-risk-inference-problem}
\end{equation} 
 An abstract inference model is shown in Figure~\ref{fig-abstract-model-inference-problem-without-complexity-constraint}.

\begin{figure}[H]
\centering
\includegraphics[width=0.7\textwidth]{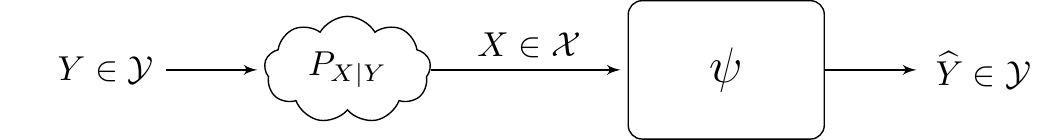}
\caption{An abstract inference model for learning.} 
\label{fig-abstract-model-inference-problem-without-complexity-constraint}
\end{figure}

The choice of a ``good'' loss function $\ell(\cdot)$ is often controversial in statistical learning theory. There is however numerical evidence that models that are trained to minimize the error's entropy often outperform ones that are trained using other criteria such as mean-square error (MSE) and higher-order statistics~\cite{E02, PEL00}. This corresponds to choosing the loss function given by the logarithmic loss, which is defined as
 \begin{equation}
\ell_{\text{log}}(y,\hat{y}) := \log \frac{1}{\hat{y}(y)}
\label{definition-per-letter-log-loss-distortion-measure}
\end{equation}
for $y \in \mc Y$, where $\hat{y} \in \mc P(\mc Y)$ designates here a probability distribution on $\mc Y$ and $\hat{y}(y)$ is the value of that distribution evaluated at the outcome $y \in \mc Y$. Although a complete and rigorous justification of the usage of the logarithmic loss as distortion measure in learning is still awaited, recently a partial explanation appeared in~\cite{PW18} where Painsky and Wornell showed that, for binary classification problems, by minimizing the logarithmic-loss one actually minimizes an upper bound to any choice of loss function that is smooth, proper (i.e., unbiased and Fisher consistent), and convex. Along the same line of work, the authors of~\cite{JCVW15} showed that under some natural data processing property Shannon's mutual information uniquely quantifies the reduction of prediction risk due to side information. Perhaps, this justifies partially why the logarithmic-loss fidelity measure is widely used in learning theory and has already been adopted in many algorithms in practice such as the \textit{infomax} criterion~\cite{L88}, the tree-based algorithm of~\cite{Q14}, or the well known Chow--Liu algorithm~\cite{CL68} for learning tree graphical models, with various applications in genetics~\cite{AMPB08}, image processing~\cite{PMV03}, computer vision~\cite{VWW97}, and others. The logarithmic loss measure also plays a central role in the theory of prediction~\cite[Ch. 09]{C-BL06}, where it is often referred to as the \textit{self-information} loss function, as well as in Bayesian modeling~\cite{LC06} where priors are usually designed     to   maximize the mutual information between the parameter to be estimated and the observations.

When the join distribution $P_{X,Y}$ is known, the optimal predictor and the minimum expected (population) risk can be characterized. Let, for every $x \in \mc X$, $\psi(x)=Q(\cdot|x) \in \mc P(\mc Y)$. It is easy to see that
\begin{subequations}
\begin{align}
\mathbb{E}_{P_{X,Y}} [\ell_{\text{log}}(Y,Q)] &= \sum_{x \in \mc X, \: y \in \mc Y} P_{X,Y}(x,y) \log\Big(\frac{1}{Q(y|x)}\Big) \\
&= \sum_{x \in \mc X, \: y \in \mc Y} P_{X,Y}(x,y) \log\Big(\frac{1}{P_{Y|X}(y|x)}\Big) + \sum_{x \in \mc X, \: y \in \mc Y} P_{X,Y}(x,y) \log\Big(\frac{P_{Y|X}(y|x)}{Q(y|x)}\Big)  \\
&= H(Y|X) + D\big(P_{Y|X}\|Q\big) \\
&\geq H(Y|X)
\end{align}
\label{optimal-decoder-no-compression}
\end{subequations}
with equality iff the predictor is given by the conditional posterior $\psi(x)=P_{Y}(Y|X=x)$. That is, the minimum expected (population) risk is given by
\begin{equation}
\min_{\psi} \mc C_{P_{X,Y}}(\psi,\ell_{\text{log}}) = H(Y|X).
\end{equation}

If the joint distribution $P_{X,Y}$ is unknown, which is most often the case in practice, the population risk as given by Equation         \eqref{definition-population-risk-inference-problem} cannot be computed directly, and, in the standard approach, one usually resorts to choosing the predictor with minimal risk on a training dataset consisting of $n$ labeled samples $\{(x_i,y_i)\}_{i=1}^n$ that are drawn independently from the unknown joint distribution $P_{X,Y}$. In this case, one is interested in optimizing the empirical population risk, which for a set of $n$ i.i.d. samples from $P_{X,Y}$,  $\mc{D}_n:=\{(x_i,y_i)\}_{i=1}^n$, is defined as
\begin{equation}
\hat{\mc C}_{P_{X,Y}}(\psi,\ell, \mc D_n ) = \frac{1}{n}\sum_{i=1}^n\ell(y_i,\psi(x_i)).
\label{definition-population-risk-inference-problem}
\end{equation} 

The difference between the empirical and population risks is normally measured in terms of the generalization gap, defined as
\begin{align}\label{eq:GenGap}
\mathrm{gen}_{P_{X,Y}}(\psi,\ell, \mc D_n):=\mc C_{P_{X,Y}}(\psi,\ell_{\text{log}})  - \hat{\mc C}_{P_{X,Y}}(\psi,\ell, \mc D_n ). 
\end{align}

\subsection{Minimum Description Length}
One popular approach to reducing  the generalization gap is by restricting the set $\mc F$ of admissible predictors to a low-complexity class (or constrained complexity) to prevent over-fitting. One way to limit the model's complexity is by restricting the range of the prediction function, as shown in Figure~\ref{fig-abstract-model-inference-problem}. This is the so-called minimum description length complexity measure, often used in the learning literature to
limit the description length of the weights of neural networks \cite{Hinton1993}. A connection between the use of the minimum
description complexity for limiting the description length of the input encoding and accuracy studied in \cite{Gilad-BachrachNT03} and with respect    to the weight complexity and accuracy
is given in \cite{AS17}. Here, the stochastic mapping $\phi \: : \: \mc X \longrightarrow \mc U$ is a compressor with
\begin{equation}
\|\phi\| \leq R
\label{inference-problem-function-range-constraint}
\end{equation}
for some prescribed ``input-complexity'' value $R$, or equivalently prescribed average description-length.

\begin{figure}[!t]
\centering
\includegraphics[width=0.95\textwidth]{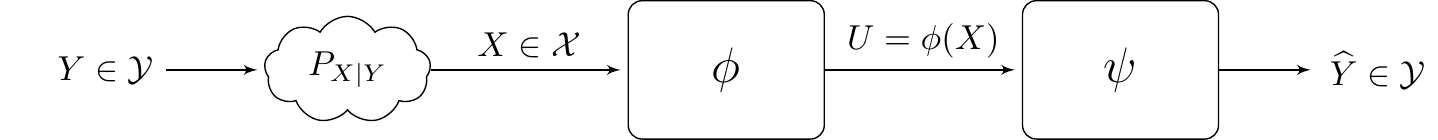}
\caption{Inference problem with constrained model's complexity.} 
\label{fig-abstract-model-inference-problem}
\end{figure}

Minimizing the constrained description length population risk is now equivalent to solving 
\begin{align}
\mc C_{P_{X,Y},\mathrm{DLC}}(R) = &\min_{\phi}\mathbb{E}_{P_{X,Y}} [\ell_{\text{log}}(Y^n,\psi(U^n))]\\
&\text{s.t. }\|\phi(X^n)\| \leq nR.
\label{population-risk-inference-problem-with-constrained-prediction-function}
\end{align}
It can be shown that this problem takes its minimum value with the choice of $\psi(U)=P_{Y|U}$ and 
\begin{equation}
\mc C_{P_{X,Y},\mathrm{DLC}}(R) = \min_{P_{U|X}} H(Y|U)\quad \text{s.t. } R\geq I(U;X),
\label{expected-logarithmic-loss-inference-problem-with-constrained-prediction-function}
\end{equation}

The solution to        \eqref{expected-logarithmic-loss-inference-problem-with-constrained-prediction-function} for different values of $R$ is effectively equivalent to the IB-problem in          \eqref{eq:IBCriteria}. Observe that the {right-hand side} of       \eqref{expected-logarithmic-loss-inference-problem-with-constrained-prediction-function} is larger for small values of $R$; it is clear that a good predictor $\phi$ should strike a right balance between reducing the model's complexity and reducing the error's entropy, or, equivalently, maximizing the mutual information $I(U;Y)$ about the target variable $Y$.

\subsection{Generalization  and Performance Bounds}

The IB-problem appears as a relevant problem in fundamental performance limits of learning. In particular, when  $P_{X,Y}$ is unknown, and instead $n$ samples i.i.d from $P_{X,Y}$ are available, the optimization of the empirical risk in  Equation         \eqref{definition-population-risk-inference-problem} leads to a mismatch between the true loss given by the population risk and the empirical risk. This gap is measured by the generalization gap in Equation~\eqref{eq:GenGap}. Interestingly, the relationship between the true loss and the empirical loss  can be bounded (in high probability) in terms of the IB-problem as \cite{vera2018role}
\begin{align*}
\mc C_{P_{X,Y}}(\psi,\ell_{\text{log}}) 
&\leq\hat{\mc C}_{P_{X,Y}}(\psi,\ell, \mc D_n ) +  \mathrm{gen}_{P_{X,Y}}(\psi,\ell, \mc D_n)  \\
&=\underbrace{H_{\hat{P}_{X,Y}^{(n)}}(Y|U)}_{\hat{\mc C}_{P_{X,Y}}(\psi,\ell, \mc D_n )}
 +\underbrace{A \sqrt{I(\hat{P}_X^{(n)};P_{U|X})}\cdot\frac{\log n}{n}+ \frac{B\sqrt{\Lambda(P_{U|X},\hat{P}_{Y|U},P_{\hat{Y}|U})}}{\sqrt{n}} +\mathcal{O}\left(\frac{\log n}{n}\right)}_{\text{Bound on }\mathrm{gen}_{P_{X,Y}}(\psi,\ell, \mc D_n) }
\end{align*}
where $\hat{P}_{U|X}$ and $\hat{P}_{Y|U}$  are the empirical encoder and decoder  and $P_{\hat{Y}|U}$ is the optimal decoder. $H_{\hat{P}_{X,Y}^{(n)}}(Y|U)$ and  $I(\hat{P}_X^{(n)};P_{U|X})$ are the empirical loss and the mutual information resulting from the dataset $\mc D_n $ and $\Lambda(P_{U|X},\hat{P}_{Y|U},P_{\hat{Y}|U})$ is a function that measures the mismatch between the optimal decoder and the empirical one.

This bound shows explicitly the trade-off between the empirical relevance and the empirical complexity. The pairs of relevance and complexity simultaneously achievable  is precisely characterized by the IB-problem. Therefore, by designing estimators based on the IB problem, as described in Section~\ref{sec:SolutionsIB}, one can perform at different regimes of performance, complexity and generalization. 

Another interesting connection between learning and the IB-method is the connection of  the logarithmic-loss as metric to common performance metrics in learning: 
\begin{itemize}
\item  The logarithmic-loss  gives an upper bound on the probability of miss-classification (accuracy):
\begin{align*}
\epsilon_{Y|X}(Q_{\hat{Y}|X}) &:= 1- \mathrm{E}_{P_{XY}}[Q_{\hat{Y}|X}]\leq 1-\exp\left( -\mathrm{E}_{P_{X,Y}}[\ell_{\text{log}}(Y,Q_{\hat{Y}|X})]  \right)
\end{align*}
\item  The logarithmic-loss is equivalent to maximum likelihood for large $n$:
\begin{align*}
-\frac{1}{n} \log P_{Y^n|X^n}(y^n|x^n) &=  -\frac{1}{n} \sum_{i=1}^n\log P_{Y|X}(y_i|x_i)\overset{n\rightarrow\infty}{\longrightarrow}  \mathrm{E}_{X,Y}[-\log P_{Y|X}(Y|X)]
\end{align*}

\item The {true distribution} $P$ minimizes the expected logarithmic-loss:
\begin{align*}
P_{Y|X}&=\arg\min_{Q_{\hat{Y}|X}} \mathrm{E}_{P} \log\frac{1}{Q_{\hat{Y}|X}} \quad \text{and }\;\min_{Q_{\hat{Y}|X}}\mathrm{E}[\ell_{\mathrm{log}}(Y,Q_{\hat{Y}|X})] =  H(Y|X)
\end{align*}

\end{itemize}

Since for $n\rightarrow \infty$ the joint distribution $P_{XY}$ can be perfectly learned, the link between these common criteria allows the use of the IB-problem to derive asymptotic performance bounds, as well as design criteria, in most of the learning scenarios of classification, regression, and inference.

\subsection{Representation Learning, ELBO and Autoencoders}
The performance of machine learning algorithms depends strongly on the choice of data representation (or features) on which they are applied. For that reason, feature engineering, i.e., the set of all pre-processing operations and transformations applied to data in the aim of making them support effective machine learning, is important. However, because it is both data- and task-dependent, such feature-engineering is labor intensive and highlights one of the major weaknesses of current learning algorithms: their inability to extract discriminative information from the data themselves  instead of hand-crafted transformations of them. In fact, although it may sometimes appear useful to deploy feature engineering in order to take advantage of human know-how and prior domain knowledge, it is highly desirable to make learning algorithms less dependent on feature engineering to make progress towards true artificial intelligence.

\textit{Representation learning} is a sub-field of learning theory that aims at learning representations of the data that make it easier to extract useful information, possibly without recourse to any feature engineering. That is, the goal is to identify and disentangle the underlying explanatory factors that are hidden in the observed data. In the case of probabilistic models, a good representation is one that captures the posterior distribution of the underlying explanatory factors for the observed input. For related works, the reader may refer, e.g., to the proceedings of the International Conference on Learning Representations (ICLR), see \url{https://iclr.cc/}.

The use of the Shannon's mutual information as a measure of similarity is particularly suitable for the purpose of learning a good representation of data \cite{huang2019universal}. 
In particular, a popular approach to representation learning are autoencoders, in which neural networks are designed for the task of representation learning. Specifically, we  design a neural network architecture such that we impose a bottleneck in the network  that  forces a compressed knowledge representation of the original input, by optimizing the Evidence Lower Bound (ELBO), given as
\begin{align}
\mathcal{L}^{\mathrm{ELBO}}({\theta, \phi, \varphi} )
:=\frac{1}{N}\sum_{i=1}^N\Big[\log Q_{\phi}(x_i| u_{i})\label{eq:ELBO}
-D_{\mathrm{KL}}(P_{\theta}(U_{i}|x_{i})\|Q_{\varphi}(U_{i}))
\Big)\Big].
\end{align}
over the neural network parameters $\theta, \phi, \varphi$.
Note that this is precisely the variational-IB cost in Equation~\eqref{eq:variational objective2} for $\beta=1$ and $Y=X$, i.e., the IB variational bound when particularized to distributions whose parameters are determined by neural networks. In addition, note that the architecture shown in Figure \ref{fig:enc_dec} is the classical neural network architecture for autoencoders, and that is coincides with the  variational IB solution resulting from the optimization of the IB-problem in Section~\ref{sssec:VIB}.
In addition, note that Equation  \eqref{eq:variational objective2}  provides an operational meaning to the $\beta$-VAE cost~\cite{HMPBGBML16}, as a criterion  to design estimators on the relevance--complexity plane for different $\beta$ values, since the $\beta$-VAE cost is given as
\begin{align}
\mathcal{L}^{\beta -\mathrm{VAE}}({\theta, \phi, \varphi} )
:=\frac{1}{N}\sum_{i=1}^N\Big[\log Q_{\phi}(x_i| u_{i})\label{eq:beta_VAE}
-\beta D_{\mathrm{KL}}(P_{\theta}(U_{i}|x_{i})\|Q_{\varphi}(U_{i}))
\Big)\Big], 
\end{align}
which coincides with the empirical version of the variational bound found in Equation \eqref{eq:variational objective2}.

\subsection{Robustness to Adversarial Attacks}
Recent advances in deep learning has allowed the design of high accuracy neural networks. However, it has been observed that the high accuracy of trained neural networks may be compromised under nearly imperceptible changes in the inputs \cite{kurakin2016adversarial, aleks2017deep, engstrom2017rotation}. 
The information bottleneck has also found applications in providing methods to improve robustness to adversarial attacks when training models.  In particular, the use of the variational IB method of Alemi et al.~\cite{AFDM17}  showed the advantages of the resulting neural network for classification in terms of robustness to adversarial attacks. Recently, alternatives strategies for extracting features in supervised learning are proposed in~\cite{pensia2019extracting} to
construct classifiers robust to small perturbations in the input space. Robustness  is measured in terms of the (statistical)-Fisher information, given for two random variables $(Y,Z)$ as
\begin{align}
\Phi(Z|Y) = \mathrm{E}_{Y,Z}\left|\frac{\partial}{\partial y }\log p(Z|Y)\right|^2.\label{eq:Fisher}
\end{align}

The method in~\cite{pensia2019extracting} builds upon the idea of the information bottleneck by introducing an additional penalty term that encourages the Fisher information in Equation         \eqref{eq:Fisher} of the extracted features to be small, when parametrized by the inputs. For this problem, under jointly Gaussian vector sources $(X,Y)$, the optimal representation is also shown to be Gaussian, in line with the results in Section~\ref{sec:Gauss} for the IB without robustness penalty. For general source distributions, a variational method is proposed similar to the variational IB method in Section~\ref{sssec:VIB}. The problem shows connections with the I-MMSE~\cite{Guo_2005}, de Brujin identity\cite{DCT91,EU14}, Cram\'er--Rao inequality~\cite{CT91}, and  Fano's inequality~\cite{CT91}.

\section{Extensions:  Distributed Information Bottleneck}\label{sec:Distributed-Information-Bottleneck}

Consider now a generalization of the IB problem in which the prediction is to be performed in a distributed manner. The model is shown in Figure~\ref{fig-abstract-model-distributed-inference}. Here, the prediction of the target variable $Y \in \mc Y$ is to be performed on the basis of samples of statistically correlated  random variables $(X_1,\hdots,X_K)$ that are observed each at a distinct predictor.
Throughout, we assume that the following Markov Chain holds for all $k \in \mc K := \{1,\hdots,K\}$,  
\begin{align}
X_{k} \mkv Y \mkv X_{\mathcal{K}/k}.
\label{eq:MKChain_pmf}
\end{align}

The variable $Y$ is a target variable and we seek to characterize how accurately  it can be predicted from a measurable random vector $(X_1,\hdots,X_K)$ when the components of this vector are  processed separately, each by a distinct encoder.

\begin{figure}[!t]
\centering
\includegraphics[width=0.8\textwidth]{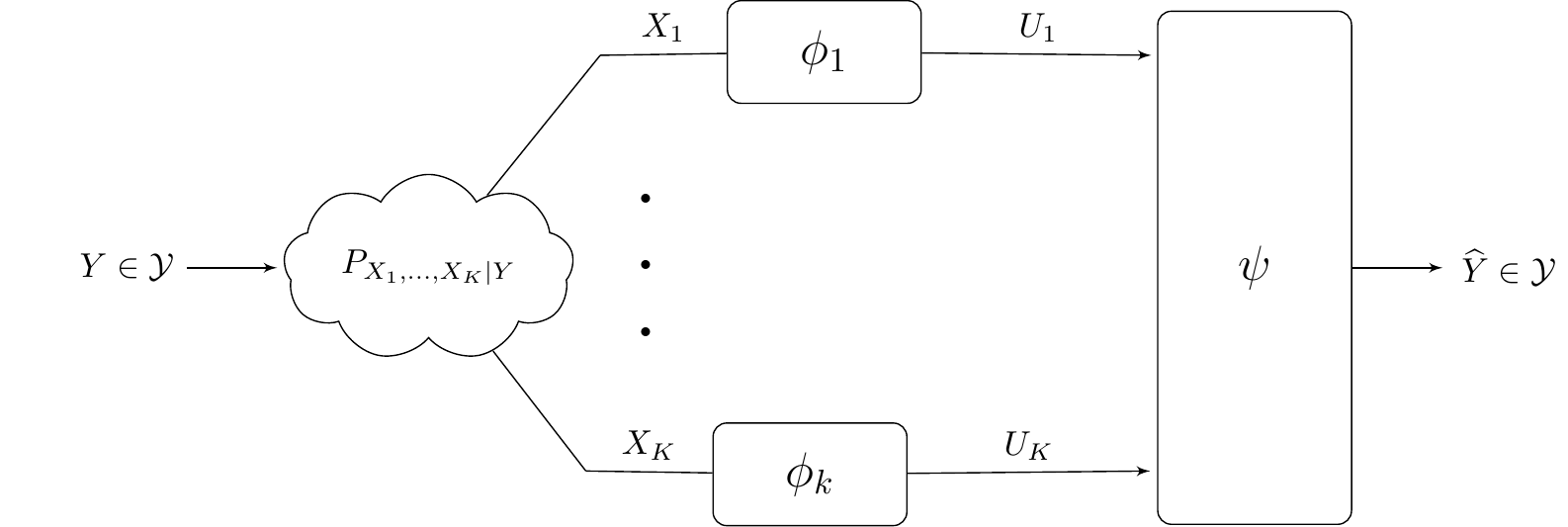}
\caption{A model for distributed, e.g., multi-view, learning.} 
\label{fig-abstract-model-distributed-inference}
\end{figure}
\subsection{The Relevance--Complexity Region}

The distributed IB problem of Figure~\ref{fig-abstract-model-distributed-inference} is studied in~\cite{E-AZ18a,E-AZ19a} from information-theoretic grounds. For both discrete memoryless (DM) and memoryless vector Gaussian models, the authors established  fundamental limits of learning in terms of optimal trade-offs between relevance and complexity, leveraging on the connection between the IB-problem and source coding. The following theorem states the result for the case of discrete memoryless sources.
\begin{new-theorem}{(\cite{E-AZ18a,E-AZ19a})}\label{theorem-relevance-complexity-region-DM-case}
The relevance--complexity region $\mathcal{IR}_{\mathrm{DIB}}$ of the distributed learning problem is given by the union of all non-negative tuples $(\Delta, R_1,\ldots, R_K)\in \mathds{R}_{+}^{K+1}$ that satisfy 
\begin{equation}
\Delta \leq \sum_{k\in \mathcal{S}} [R_k\!-\!I(X_{k};U_{k}|Y,T)]  + I(Y;U_{\mathcal{S}^c}|T), \quad \forall \mc S \subseteq \mc K
\label{eq:ComplexityrelevanceFunction}
\end{equation}
for some joint distribution of the form $P_T P_Y \prod_{k=1}^K P_{X_k|Y} \prod_{k=1}^{K}P_{U_k|X_k,T}$.
\end{new-theorem}

\begin{proof}
The proof of Theorem~\ref{theorem-relevance-complexity-region-DM-case} can be found in~\cite[Section 7.1]{E-AZ19a}  and is reproduced in Section~\ref{appendix-proof-theorem-relevance-complexity-region-DM-case} for completeness.
\end{proof}

For a given joint data distribution $P_{X_{\mc K},Y}$, Theorem \ref{theorem-relevance-complexity-region-DM-case} extends the single encoder IB principle of
Tishby in Equation         \eqref{IB-Lagrangian-formulation-minimizing-complexity} to the distributed learning model with $K$ encoders, which we denote by Distributed Information Bottleneck (DIB) problem. The result characterizes the optimal relevance--complexity trade-off as a region of achievable
tuples $(\Delta, R_1, \ldots, R_K)$ in terms of a distributed representation learning problem involving the optimization over $K$
conditional pmfs $P_{U_k|X_k,T}$ and a pmf $P_T$ . The pmfs $P_{U_k|X_k,T}$ correspond to stochastic encodings of the observation
$X_k$ to a latent variable, or representation, $U_k$ which captures the relevant information of $Y$ in observation $X_k$.
Variable $T$ corresponds to a time-sharing among different encoding mappings  (see, e.g., \cite{GK11}). For such encoders,
the optimal decoder is implicitly given by the conditional pmf of $Y$ from $U_1, \ldots,   U_K$, i.e., $P_{Y|U_{\mc K},T}$.

The characterization of the relevance--complexity region can be used to derive a cost function for the D-IB similarly to the IB-Lagrangian in Equation         \eqref{IB-Lagrangian-formulation-minimizing-complexity}. For simplicity, let us consider the problem of maximizing the relevance under a sum-complexity constraint. Let $R_{\mathrm{sum}} =\sum_{k=1}^K R_k$ and
\begin{align}
\mc{RI}_{\mathrm{DIB}}^{\mathrm{sum}}: = &\Big\{(\Delta, R_{\mathrm{sum}})\in \mathds{R}_+^2:\exists (R_1,\ldots, R_K)\in \mathds{R}_+^K\text{ s.t. }\nonumber
 \sum_{k=1}^K R_k = R_{\mathrm{sum}} \:\: \text{and}\:\: (\Delta, R_1,\ldots, R_K)\in \mc{RI}_{\mathrm{DIB}} \Big\}. 
\end{align}

We define the DIB-Lagrangian (under sum-rate) as
\begin{align}
\mc L_{s}(\dv P) :=- H(Y|U_{\mc K}) - s \sum_{k=1}^K [H(Y|U_k) +I(X_k;U_k) ].
\label{eq:CostF}
\end{align}

The optimization of Equation         \eqref{eq:CostF} over the encoders $P_{U_k|X_k,T}$ allows   obtaining  mappings that perform on the boundary of the relevance--sum complexity region $\mc{RI}_{\mathrm{DIB}}^{\mathrm{sum}}$. 
To see that, note that it is easy to see that the relevance--sum complexity region $\mc{RI}_{\mathrm{DIB}}^{\mathrm{sum}}$ is composed of all the pairs $(\Delta, R_{\mathrm{sum}})\in \mathds{R}_{+}^{2}$ for which $\Delta \leq \Delta(R_{\mathrm{sum}},P_{X_{\mc K},Y})$, with
\begin{equation}
\Delta(R_{\mathrm{sum}},P_{X_{\mc K},Y}) = \max_{\dv P}  \min\left\{I(Y; U_{\mc K}),R_{\mathrm{sum}}- \sum_{k=1}^KI(X_k;U_k|Y)\right\}, 
\label{eq:RelevanceSumComplexityFunction}
\end{equation} 
where the maximization is over joint distributions that factorize as $P_Y \prod_{k=1}^K P_{X_k|Y} \prod_{k=1}^{K}P_{U_k|X_k}$. The pairs $(\Delta,R_{\mathrm{sum}})$ that lie on the boundary of $\mc {RI}^{\mathrm{sum}}_{\mathrm{DIB}}$ can be characterized as given in the following proposition. 

\begin{new-proposition}\label{proposition-parametrization-relevance-complexity-region-DM-case}
For every pair $(\Delta,R_{\mathrm{sum}})\in \mathds{R}^2_+$ that lies on the boundary of the region $\mc {RI}^{\mathrm{sum}}_{\mathrm{DIB}}$, there exists a parameter $s \geq 0$ such that $(\Delta,R_{\mathrm{sum}}) = (\Delta_{s}, R_{s})$, with
\begin{align}
&\Delta_{s} =\frac{1}{(1+s)} \left[(1+sK)H(Y)  +  s R_{s}+ \max_{\dv P}\mc L_{s}(\dv P)\right],\label{eq:Dparam}\\
&R_{s} = I(Y;U_{\mc K}^*) + \sum_{k=1}^K [I(X_k;U_k^*) - I(Y;U_k^*)],\label{eq:R1param}
\end{align}
where $\dv P^*$ is the set of conditional pmfs $\dv P = \{P_{U_1|X_1},\hdots, P_{U_K|X_K}\}$ that maximize the cost function  in  Equation~\eqref{eq:CostF}
\end{new-proposition}

\begin{proof}
The proof of Proposition~\ref{proposition-parametrization-relevance-complexity-region-DM-case} can be found in~\cite[Section 7.3]{E-AZ19a}  and is reproduced here in Section~\ref{appendix-proof-proposition-parametrization-relevance-complexity-region-DM-case} for completeness.
\end{proof}

The optimization of the distributed IB cost function  in  Equation         \eqref{eq:CostF} generalizes the centralized Tishby's information bottleneck formulation in Equation         \eqref{IB-Lagrangian-formulation-minimizing-complexity} to the distributed learning setting. Note that for $K = 1$ the optimization in Equation         \eqref{eq:Dparam} reduces to the single encoder cost in Equation         \eqref{IB-Lagrangian-formulation-minimizing-complexity} with a multiplier $s/(1+s)$.

\subsection{Solutions to the Distributed Information Bottleneck}
The methods described in Section~\ref{sec:SolutionsIB} can be extended to the distributed information bottleneck case in order to find the mappings $P_{U_1|X_1,T},\cdots, P_{U_K|X_K,T} $ in different scenarios. 

\subsubsection{Vector Gaussian Model}\label{sec:Gauss}

In this section, we show that for the jointly vector Gaussian data model it is enough to restrict to Gaussian auxiliaries $(\dv U_1,\hdots,\dv U_K)$ in order to exhaust the entire relevance--complexity region. In addition, we provide an explicit analytical expression of this region. Let $(\mathbf{X}_1,\ldots,\mathbf{X}_K,\mathbf{Y})$ be a jointly vector Gaussian vector that satisfies the Markov Chain in   Equation         \eqref{eq:MKChain_pmf}. Without loss of generality, let the target variable be a complex-valued, zero-mean multivariate Gaussian $\mathbf{Y}\in \mathds{C}^{n_y}$ with  covariance matrix $\mathbf{\Sigma}_{\dv y}$, i.e., $\mathbf{Y}\sim \mc{CN}(\dv y; \dv 0, \mathbf{\Sigma}_{\dv y})$, and $\dv X_{k}\in \mathds{C}^{n_k}$ given by
\begin{equation}
\mathbf{X}_k = \mathbf{H}_{k}\mathbf{Y}+\mathbf{N}_k,
\label{mimo-gaussian-model}
\end{equation}
where $\mathbf{H}_{k}\in \mathds{C}^{n_k\times n_y}$ models the linear  model connecting $\dv Y$ to the observation at encoder $k$  and $\mathbf{N}_k\in\mathds{C}^{n_k}$ is the noise vector at encoder $k$, assumed to be Gaussian with zero-mean, covariance matrix $\mathbf{\Sigma}_{k}$, and independent from all other noises and $\dv Y$. 
 
For the vector Gaussian model Equation         \eqref{mimo-gaussian-model}, the result of Theorem~\ref{theorem-relevance-complexity-region-DM-case}, which can be extended  to continuous sources using standard techniques, characterizes the relevance--complexity region of this model. The following theorem characterizes the relevance--complexity region, which we denote hereafter as $\mc {RI}_{\mathrm{DIB}}^{\mathrm{G}}$. The theorem also shows that in order to exhaust this region it is enough to restrict to no time sharing, i.e., $T=\emptyset$  and multivariate Gaussian test channels
\begin{equation}
\dv U_k = \dv A_k \dv X_k + \dv Z_k\sim \mc {CN}(\dv u_k;\dv A_k \dv X_k, \dv \Sigma_{z,k}),
\label{eq:GaussMap}
\end{equation}
where $\dv A_k \in \mathds{C}^{n_k\times n_k}$ projects  $\dv X_k$ and $\dv Z_k$ is a zero-mean Gaussian noise with covariance $\dv \Sigma_{z,k}$. 

\begin{new-theorem}\label{theorem-relevance-complexity-region-vector-Gaussian-case}
For the vector Gaussian data model, the relevance--complexity region $\mc {RI}_{\mathrm{DIB}}^{\mathrm{G}}$  is given by the union of all tuples $(\Delta, R_1,\ldots,R_L)$ that satisfy 
\begin{equation*}
\Delta \leq \sum_{k\in\mathcal{S}}\left(R_k+\log\left|\dv I-\mathbf{\Sigma}_{k}^{1/2}\mathbf{\Omega}_{k}\mathbf{\Sigma}_{k}^{1/2}\right|\right) + \log\left| \mathbf{I} +\sum_{k\in\mathcal{S}^{c}}\mathbf{\Sigma}_{\dv y}^{1/2}\mathbf{H}_{k}^{\dagger}
\mathbf{\Omega}_{k}
\mathbf{H}_{k}\mathbf{\Sigma}_{\dv y}^{1/2}\right|, \quad \forall \mc S \subseteq \mc K,
\end{equation*}
for some  matrices $\dv 0 \preceq\dv \Omega_k\preceq\dv \Sigma_k^{-1}$. 
\end{new-theorem}

\begin{proof}
The proof of Theorem~\ref{theorem-relevance-complexity-region-vector-Gaussian-case} can be found in~\cite[Section 7.5]{E-AZ19a}  and is reproduced here in Section~\ref{appendix-proof-theorem-relevance-complexity-region-vector-Gaussian-case} for completeness.
\end{proof}

Theorem~\ref{theorem-relevance-complexity-region-vector-Gaussian-case} extends the result of~\cite{CGTW05, WFM14} on the relevance--complexity trade-off characterization of the single-encoder IB problem for jointly Gaussian sources to $K$ encoders. The theorem also shows that the optimal test channels $P_{U_k|X_k}$ are multivariate Gaussian, as given by Equation         \eqref{eq:GaussMap}. 

Consider the following symmetric distributed scalar Gaussian  setting, in which  $Y \sim \mc N(0,1)$ and
\begin{subequations}
\begin{align}
X_1 &= \sqrt{\text{snr}} Y + N_1 \\
X_2 &= \sqrt{\text{snr}} Y + N_2
\end{align}
\end{subequations}
where $N_1$ and $N_2$ are standard Gaussian with zero-mean and unit variance, both independent of $Y$. In this case, for $I(U_1;X_1)=R$ and $I(U;X_2)=R$, the optimal relevance is 
\begin{equation}
\Delta^{\star}(R, \text{snr}) = \frac{1}{2} \log \Big(1+ 2 \text{snr} \exp(-4R) \Big(\exp(4R) + \text{snr} - \sqrt{\text{snr}^2 + (1+2\text{snr})\exp(4R)} \Big)\Big). 
\end{equation}

 An easy upper bound on the relevance can be obtained by assuming that $X_1$ and $X_2$ are encoded jointly at rate $2R$, to get
\begin{equation}
\Delta_{\text{ub}}(R,\text{snr}) = \frac{1}{2} \log(1+2\text{snr}) - \frac{1}{2}\log\Big(1 + 2 \text{snr} \exp(-4R)\Big).
\end{equation} 

 Note that, if $X_1$ and $X_2$ are encoded independently, an achievable relevance level is given by
\begin{equation}
\Delta_{\text{lb}}(R,\text{snr}) = \frac{1}{2} \log(1+2\text{snr} - \text{snr} \exp(-2R)) - \frac{1}{2}\log\Big(1 + \text{snr} \exp(-2R)\Big).
\end{equation}

\subsection{Solutions for Generic Distributions}
Next, we present how the distributed information bottleneck can be solved for generic distributions. Similar  to the case of single encoder IB-problem, the solutions are based on a variational bound on the DIB-Lagrangian. For simplicity, we look at the D-IB under sum-rate constraint~\cite{E-AZ19a}.
\subsection{A Variational Bound}

The optimization of Equation         \eqref{eq:CostF} generally requires   computing  marginal distributions that involve the descriptions $U_1,\hdots, U_K$, which might not be possible in practice. In what follows, we derive a variational lower bound on $\mc L_s(\dv P)$ on the DIB cost function in terms of families of stochastic mappings $Q_{Y|U_{1},\dots, U_K}$ (a decoder), $\{Q_{Y|U_k}\}_{k=1}^K$ and priors $\{Q_{U_k}\}_{k=1}^K$. For the simplicity of the notation, we let
\begin{equation}
\dv Q := \{Q_{Y|U_1,\ldots, U_K},Q_{Y|U_1},\ldots, Q_{Y|U_K},Q_{U_1},\ldots, Q_{U_K}\}.
\end{equation}

The variational D-IB cost for the DIB-problem is given by
\begin{equation}
\mc L^{\mathrm{VB}}_s(\dv P, \dv Q) :=   \underbrace{\mathds{E}[\log Q_{Y|U_{\mc K}}(Y| U_{\mc K})]}_{\text{av. logarithmic-loss}} + s \underbrace{\sum_{k=1}^K\Big( \mathds{E}[\log  Q_{Y|U_k}(Y|U_k)]- D_{\mathrm{KL}}(P_{U_k|X_k}\| Q_{U_k}) \Big)}_{\text{regularizer}}.
\label{eq:FunctionPQ}
\end{equation}

\begin{new-lemma}\label{lemma:QUpdate}
For fixed $\dv P$, we have
\begin{align}
\mc L_s(\dv P) \geq \mc L^{\mathrm{VB}}_s(\dv P, \dv Q), \qquad \text{for all pmfs } \dv Q.
\end{align}
In addition, there exists a unique $\dv Q$ that achieves the maximum $\max_{\dv Q}\mc L^{\mathrm{VB}}_s(\dv P, \dv Q) = \mc L_s(\dv P)$, and is given by, $\forall k \in \mc K$,
\begin{subequations}
\begin{align}
\label{eq:Qstark}
Q^*_{U_k} &= P_{U_k} \\
Q^*_{Y|U_k} &= P_{Y|U_k} \\
Q^*_{Y|U_1,\ldots,U_k} &= P_{Y|U_1,\ldots, U_K}, 
\end{align}
\label{eq:Qstarall}
\end{subequations}
where the marginals $P_{U_k}$ and the conditional marginals $P_{Y|U_k}$ and $P_{Y|U_1,\ldots, U_K}$ are computed from $\dv P$.
\end{new-lemma}

\begin{proof}
The proof of Lemma~\ref{lemma:QUpdate} can be found in~\cite[Section 7.4]{E-AZ19a} and is reproduced here in Section~\ref{appendix-proof-lemma:QUpdate} for completeness.
\end{proof}

Then, the optimization in Equation         \eqref{eq:Dparam} can be written in terms of the variational DIB cost function as follows,
\begin{equation}
\max_{\dv P}\mc L_{s}(\dv P) = \max_{\dv P}\max_{\dv Q}\mc L^{\mathrm{VB}}_{s}(\dv P,\dv Q).
\label{eq:VarEq}
\end{equation}

\noindent The  variational DIB  cost in Equation         \eqref{eq:FunctionPQ} is a generalization to distributed learning with $K$-encoders of the evidence lower bound (ELBO) of the target variable $Y$ given the representations $U_1,\hdots, U_K$~\cite{KW13}. If $Y =(X_1,\ldots, X_K)$, the bound generalizes the ELBO used for VAEs to the setting of $K\geq 2$ encoders. In addition, note that Equation         \eqref{eq:FunctionPQ} also generalizes and provides an operational meaning to the $\beta$-VAE cost~\cite{HMPBGBML16} with $\beta = s/(1+s)$, as a criteria to design estimators on the relevance--complexity plane for different $\beta$ values.

\subsection{Known Memoryless Distributions}
When the data model is discrete and the joint distribution $P_{X,Y}$ is  known, the DIB problem can be solved by using an
 iterative method that optimizes the variational IB cost function in Equation         \eqref{eq:VarEq}
alternating over the distributions $\dv P,\dv Q$.
The optimal {encoders and decoders of the D-IB under sum-rate constraint satisfy the following self consistent equations,
\begin{align*}
p(u_k|y_k) &= \frac{p(u_k)}{Z(\beta,u_k)}\exp\left(-\psi_s(u_k,y_k)\right),\\
p(x|u_k) &=\sum_{y_k\in \mc Y_k} p(y_k|u_k)p(x|y_k)\\
p(x|u_1,\ldots,u_K) &=\sum_{y_{\mc K}\in \mc Y_{\mc K}}p(y_{\mc K}) p(u_{\mc K}|y_{\mc K})p(x|y_{\mc K})/p(u_{\mc K})
\end{align*}
where
$\psi_s(u_k,y_k):= D_{\mathrm{KL}}(P_{X|y_k}||Q_{X|u_k})
+\frac{1}{s}
\mathrm{E}_{U_{\mc K\setminus k}|y_k}[D_{\mathrm{KL}}(P_{X|U_{\mc K\setminus k},y_k}||Q_{X|U_{\mc K\setminus k},u_k}))]$.

Alternating iterations of these equations converge to a  solution for any initial $p(u_k|x_k)$, similarly to a {Blahut--Arimoto }algorithm and the EM. 

\subsubsection{Distributed Variational IB}

When the data distribution is unknown and only data samples are available, the variational DIB cost in Equation         \eqref{eq:VarEq}  can be optimized following similar steps as for the variational IB in Section~\ref{sssec:VIB} by parameterizing the encoding and decoding distributions  $\dv P,\dv Q$  using a family of distributions whose parameters are determined by DNNs. This
allows us to formulate Equation         \eqref{eq:VarEq}  in terms of the DNN parameters, i.e., its weights, and optimize it by using the reparameterization
trick~\cite{KW13}, Monte Carlo sampling, and stochastic gradient descent (SGD)-type algorithms.

\begin{figure}[!t]
\centering
\includegraphics[width=0.73\textwidth]{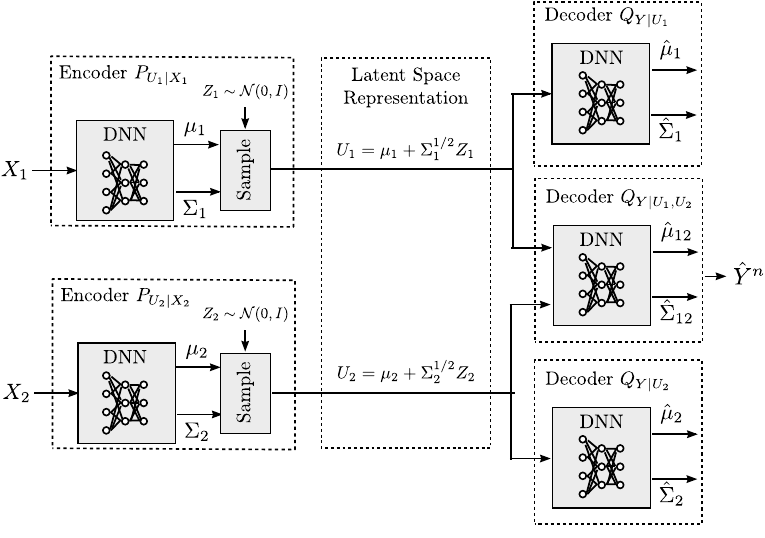}
\caption{Example parameterization of the {Distributed Variational Information B}ottleneck method using neural networks.}
\label{fig:latentmodel}
\end{figure}

Considering encoders and decoders $\dv P,\dv Q$ parameterized by  DNN parameters  $\boldsymbol{\theta},\boldsymbol{\phi},\boldsymbol\varphi$, the DIB cost in Equation         \eqref{eq:VarEq} can be optimized by considering the following empirical Monte Carlo approximation:
\begin{align} 
 \max_{\boldsymbol{\theta},\boldsymbol{\phi}, \boldsymbol\varphi}\frac{1}{n}\sum_{i=1}^n\Big[\!\log Q_{\phi_{\mc K}}(y_i| u_{1,i,j},\ldots, u_{K,i,j} )
+ s\sum_{k=1}^K \!\Big(\! \log Q_{\phi_{k}}(y_i|u_{k,i,j})\!-\!D_{\mathrm{KL}}(P_{\theta_k}(U_{k,i}|x_{k,i})\|Q_{\varphi_k}(U_{k,i}))
\Big)\Big],\label{eq:Lhatoptimization}
\end{align}
where  $u_{k,i,j} = g_{\phi_k}(x_{k,i},z_{k,j})$ are samples obtained from the reparametrization trick by sampling  from $K$ random variables $P_{Z_k}$. The details of the method can be found in~\cite{E-AZ19a}. The resulting architecture is shown in Figure~\ref{fig:latentmodel}. This architecture generalizes that from autoencoders to the distributed setup with $K$ encoders.

\subsection{Connections to Coding Problems and Learning}
Similar to the point-to-point IB-problem, the distributed IB problem also has abundant connections with (asymptotic) coding and learning problems. 

\subsubsection{Distributed Source Coding under Logarithmic Loss}

Key element to the proof of the converse part of Theorem~\ref{th:MK_C_Main} is the connection with the Chief Executive Officer (CEO) source coding problem. For the case of $K \geq 2$ encoders, while the characterization of the optimal rate-distortion region of this problem for general distortion measures has eluded the information theory for now more than four decades, a characterization of the optimal region in the case of logarithmic loss distortion measure has been provided recently in~\cite{CW14}. A key step in~\cite{CW14} is that the log-loss distortion measure admits a lower bound in the form of the entropy of the source conditioned on the decoders' input. Leveraging   this result, in our converse proof of Theorem~\ref{th:MK_C_Main}, we derive a single letter  upper bound on the entropy of the channel inputs conditioned on the indices $J_{\mc K}$ that are sent by the relays, in the absence of knowledge of the codebooks indices $F_{\mc L}$.  In addition, the rate region of the vector Gaussian CEO problem under logarithmic loss distortion measure has been found recently in~\cite{UEZa,UEZb}.

\subsubsection{Cloud RAN}

Consider the discrete memoryless (DM) CRAN model shown in Figure~\ref{fig:Schm}. In this model, $L$ users communicate with a common destination or central processor (CP) through $K$ relay nodes, where $L \geq 1$ and $K \geq 1$. Relay node $k$, $ 1 \leq k \leq K$, is connected to the CP via an error-free finite-rate fronthaul link of capacity $C_k$. In what follows, we let $\mc L := [1\!:\!L]$ and $\mc K := [1\!:\!K]$ indicate the set of users and relays, respectively. Similar to~\cite{SES11}, the relay nodes are constrained to operate without knowledge of the users' codebooks and only know a time-sharing sequence $Q^n$, i.e., a set of time instants at which users switch among different codebooks. The obliviousness of the relay nodes to the actual codebooks of the users is modeled via the notion of \textit{randomized encoding} \cite{SSSK08,LN98}. That is, users or transmitters select their codebooks at random and the relay nodes are \textit{not} informed about the currently selected  codebooks, while the CP is given such information.

\begin{figure}[!t]
\centering
\includegraphics[width=0.85\textwidth]{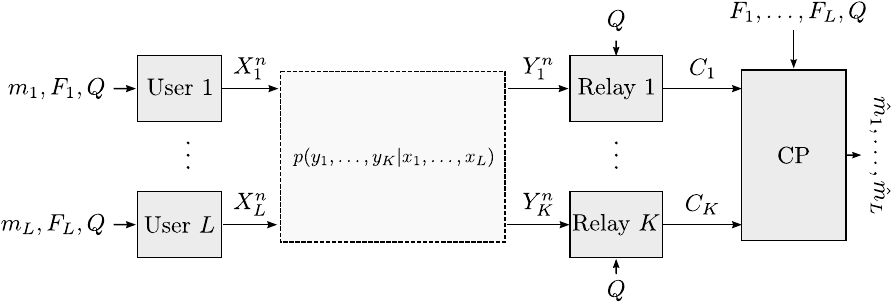}
\caption{CRAN model with oblivious relaying and time-sharing.} 
\label{fig:Schm}
\end{figure}


 Consider the following class of DM CRANs in which the channel outputs at the relay nodes are independent conditionally on the users' inputs. That is, for all $k \in \mc K$ and all $i \in [1\!:\!n]$, 
\begin{align}
Y_{k,i} \mkv X_{\mathcal{L},i} \mkv Y_{\mathcal{K}/k,i}
\label{eq:MKChain_pmf}
\end{align}  
forms a Markov Chain in this order. 

The following theorem provides a characterization of the capacity region of this class of DM CRAN problem under oblivious relaying.
\begin{new-theorem}[\cite{E-AZCS17a, E-AZCS19a}]~\label{th:MK_C_Main}
For the class of DM CRANs with oblivious relay processing and enabled time-sharing for which Equation         \eqref{eq:MKChain_pmf} holds, the capacity region $\mc C(C_{\mc K})$ is given by the union of all rate tuples $(R_1,\ldots, R_L)$ which satisfy 
\begin{align*}
\sum_{t\in \mathcal{T}}R_t\leq& \sum_{s\in \mathcal{S}} [C_s-I(Y_{s};U_{s}|X_{\mathcal{L}},Q)] 
+ I(X_{\mathcal{T}};U_{\mathcal{S}^c}|X_{\mathcal{T}^c},Q),
\end{align*}
for all non-empty subsets $\mathcal{T} \subseteq \mathcal{L}$ and all $\mathcal{S} \subseteq \mathcal{K}$, for some joint measure of the form 
\begin{align}\label{eq:PMF_Cap}
p(q)\prod_{l=1}^L p(x_l|q)\prod_{k=1}^Kp(y_k|x_{\mathcal{L}})\prod_{k=1}^{K}p(u_k|y_k,q).
\end{align}
\end{new-theorem}
The direct part of Theorem~\ref{th:MK_C_Main} can be obtained by a coding scheme in which each relay node compresses its channel output by using Wyner--Ziv binning to exploit the correlation with the channel outputs at the other relays, and forwards the bin index to the CP over its rate-limited link. The CP jointly decodes the compression indices (within the corresponding bins) and the transmitted messages, i.e., Cover-El Gamal compress-and-forward~\cite[Theorem 3]{CG79} with joint decompression and decoding (CF-JD). Alternatively, the rate region of Theorem~\ref{th:MK_C_Main} can also be obtained by a direct application of the noisy network coding (NNC) scheme of~\cite[Theorem 1]{H-LKGC11}. 

The connection between this problem, source coding and the distributed  information bottleneck is discussed in \cite{E-AZCS17a, E-AZCS19a}, particularly in the derivation of the converse part of the theorem. Note also  the similarity between the resulting capacity region in Theorem~\ref{th:MK_C_Main}  and the relevance complexity region of the distributed information bottleneck in Theorem~
\ref{theorem-relevance-complexity-region-DM-case}, despite the significant differences of the setups.

\subsubsection{Distributed Inference, ELBO and Multi-View Learning}

In many data analytics problems, data  are collected from various sources of information or feature extractors  and are  intrinsically \textit{heterogeneous}. For example, an image can be identified by its color or texture features  and a document may contain text and images. Conventional machine learning approaches concatenate all available data into one big row vector (or matrix) on which a suitable algorithm is then applied. Treating different observations as a single source might cause overfitting and is not physically meaningful because each group of data may have different statistical properties. Alternatively, one may partition the data into groups according to samples homogeneity, and each group of data be regarded as a separate \textit{view}. This paradigm, termed \textit{multi-view learning}~\cite{XTX13}, has received growing interest, and various algorithms exist, sometimes under references such as \textit{co-training}~\cite{BM98,DFU11,KD11,GA11}, \textit{multiple kernel learning}~\cite{GA11}, and \textit{subspace learning}~\cite{JSD10}. By using distinct encoder mappings to represent distinct groups of data, and jointly optimizing over all mappings to remove redundancy, multi-view learning offers a degree of flexibility that is not only desirable in practice but is also likely to result in better learning capability. Actually, as shown in~\cite{V13}, local learning algorithms produce fewer errors than global ones. Viewing the problem as that of function approximation, the intuition is that it is usually not easy  to find a unique function that holds good predictability properties in the entire data space. 

Besides, the distributed learning of Figure~\ref{fig-abstract-model-distributed-inference} clearly finds application in all those scenarios in which learning is performed collaboratively but distinct learners either only access subsets of the entire dataset (e.g., due to physical constraints) or   access independent noisy versions of the entire dataset. 

In addition, similar  to the single encoder case, the distributed IB also finds applications in fundamental performance limits and formulation of cost functions from an operational point of view. One of such examples is the generalization of the commonly used ELBO and given in Equation         \eqref{eq:ELBO} to the setup with $K$ views or observations, as formulated in Equation         \eqref{eq:FunctionPQ}. Similarly, from the formulation of the DIB problem, a natural generalization of the classical autoencoders emerge, as given in Figure~\ref{fig:latentmodel}.

\section{Outlook}

A variant of the bottleneck problem in which the encoder's output is constrained in terms of its entropy, rather than its mutual information with the encoder's input as done originally in~\cite{TPB99}, was considered in~\cite{SS16}. The solution of this problem turns out to be a deterministic encoder map as opposed to the stochastic encoder map that is optimal under the IB framework of~\cite{TPB99}, which results in a reduction of the algorithm's complexity. This idea was then used and extended to the case of available resource (or time) sharing in~\cite{HPS18}. 

In the context of privacy against inference attacks~\cite{P-CF12}, the authors of~\cite{MSFM14,ADAL19} considered a dual of the information bottleneck problem in which $X \in \mc X$ represents some private data that  are correlated with the non-private data $Y \in \mc Y$. A legitimate receiver (analyst) wishes to infer as much information as possible about the non-private data $Y$ but does not need to infer any information about the private data $X$. Because $X$ and $Y$ are correlated, sharing the non-private data $X$ with the analyst possibly reveals information about $Y$. For this reason, there is a trade-off between the amount of information that the user shares about $X$ as measured by the mutual information $I(U;X)$ and the information that he keeps private about $Y$ as measured by the mutual information $I(U;Y)$, where $U=\phi(X)$.

 Among interesting problems that are left unaddressed in this paper is that of characterizing optimal input distributions under rate-constrained compression at the relays where, e.g., discrete signaling is already known to sometimes outperform Gaussian signaling for single-user Gaussian CRAN~\cite{SSSK08}. It is conjectured that the optimal input distribution is discrete. Other issues might relate to extensions to continuous time filtered Gaussian channels, in parallel to the regular bottleneck problem~\cite{HPS18}, or extensions to settings in which fronthauls may be not available at some radio-units, and that is unknown to the systems. That is, the more radio units are connected to the central unit, the higher is the rate that  could be conveyed over the CRAN uplink~\cite{KSS13}. Alternatively, one may consider finding the worst-case noise under given input distributions, e.g., Gaussian, and rate-constrained compression at the relays. Furthermore, there are interesting aspects that address processing constraints of continuous waveforms, e.g.,  sampling at a given rate~\cite{CGE14,KEG18} with focus on remote logarithmic distortion~\cite{CW14}, which in turn boils down to the distributed bottleneck problem~\cite{E-AZ18a,E-AZ19a}. We also mention finite-sample size analysis (i.e., finite block length $n$, which relates to the literature on finite block length coding in information theory). Finally, it is interesting to observe that the bottleneck problem relates to interesting problem when $R$ is not necessarily scaled with the block length $n$.

\section{Proofs}

\subsection{Proof of Theorem~\ref{theorem-relevance-complexity-region-DM-case}}\label{appendix-proof-theorem-relevance-complexity-region-DM-case}

The proof relies on the equivalence of the studied distributed learning problem with the Chief-Executive Officer (CEO) problem under logarithmic-loss distortion measure, which was studied {in~}\cite[Theorem 10]{CW14}. For the $K$-encoder CEO problem, let us consider $K$ encoding functions  $\phi_k:\mc X_k\rightarrow \mc M_k^{(n)}$ satisfying $nR_k \geq \log |\phi_k(X^n_k)|$  and a decoding function $\tilde{\psi}: \mathcal{M}^{(n)}_1\times \hdots \times \mathcal{M}^{(n)}_K \rightarrow \mathcal{\hat{Y}}^n$,  which produces a probabilistic estimate of $Y$ from the outputs of the encoders, i.e., $\mathcal{\hat{Y}}^n$ is the set of distributions on $\mc Y$. The quality of the estimation is measured in terms of the average log-loss. 

\begin{new-definition}
A tuple $(D,R_1,\ldots, R_K)$ is said to be achievable in the $K$-encoder CEO problem for $P_{X_{\mc K},Y}$ for which the Markov Chain in   Equation         \eqref{eq:MKChain_pmf} holds, if 
there exists a length $n$, encoders $\phi_k$ for $k\in \mc K$, and a decoder $\tilde{\psi}$, 
such that
\begin{align}
D & \geq  \mathrm{E}\left[\frac{1}{n}\log\frac{1}{\hat{P}_{Y^n|J_{\mc K}}(Y^n|\phi_1(X_1^n),\hdots, \phi_K(X_K^n))}\right], \\
R_k &\geq \frac{1}{n}\log |\phi_k(X_k^n)| \quad \text{for all}\:\:\: k \in \mc K.
\end{align}
{The rate}-distortion region $\mathcal{RD}_{\mathrm{CEO}}$ is given by the closure of all achievable tuples $(D,R_1,\ldots, R_K)$. \qed
\end{new-definition} 

 The following lemma shows that the minimum average logarithmic loss is the conditional entropy of $Y$ given the descriptions. The result is essentially equivalent to \cite[Lemma 1]{CW14} and it is provided for completeness.
\begin{new-lemma}\label{lem:LL_Relevance}
Let us consider $P_{X_{\mc K}, Y}$ and the encoders $J_k = \phi_k(X_k^n)$, $k\in \mc K$ and the decoder $\hat{Y}^n = \tilde{\psi}(J_{\mc K})$. Then,
\begin{align}
\mathrm{E}[\ell_{\mathrm{log}}(Y^n,\hat{Y}^n)]\geq H(Y^n|J_{\mc K}),
\end{align} 
with equality if and only if $\tilde{\psi}(J_{\mc K}) = \{P_{Y^n|J_{\mc K}}(y^n|J_{\mc K})\}_{y^n\in \mc Y^n}$. 
\end{new-lemma}
\begin{proof}
Let $Z := (J_1,\ldots, J_K)$ be the argument of $\tilde{\psi}$ and $\hat{P}(y^n|z)$ be a distribution on $\mc{Y}^n$. We have for $Z=z$:
\begin{align}
\mathrm{E}[\ell_{\mathrm{log}}(Y^n,\hat{Y}^n)|Z=z]
&= \sum_{y^n\in\mc{Y}^n}P(y^n|z)\log\left(\frac{1}{\hat{P}(y^n|z)}\right)\\
&= \sum_{y^n\in\mc{Y}^n}P(y^n|z)\log\left(\frac{P(y^n|z)}{\hat{P}(y^n|z)}\right) + H(Y^n|Z=z)\\
&= D_{\mathrm{KL}}(P(y^n|z)\|\hat{P}(y^n|z))+ H(Y^n|Z=z)\\
&\geq H(Y^n|Z=z),\label{eq:LowerBound}
\end{align}
where Equation         \eqref{eq:LowerBound} is due to the non-negativity of the KL divergence and the equality holds if and only if for $\hat{P}(y^n|z) = P(y^n|z)$ where $ P(y^n|z) = \mathrm{Pr}\{Y^n=y^n|Z=z\}$ for all $z$ and $y^n\in \mc Y^n$. Averaging over  $Z$ completes the proof.
\end{proof}

 Essentially, Lemma~\ref{lem:LL_Relevance} states that  minimizing the average log-loss is equivalent to maximizing relevance as given by the mutual information $I\Big(Y^n; \psi\Big(\phi_1(X^n_1),\hdots,\phi_K(X^n_K)\Big)\Big)$. Formally, the connection between the distributed learning problem under study and the $K$-encoder CEO problem studied in~\cite{CW14} can be formulated as stated next.
\begin{new-proposition}\label{prop:Eqreg}
A tuple $(\Delta,R_1,\ldots, R_K)\in \mc {RI}_{\mathrm{DIB}}$ if and only if $(H(Y)-\Delta,R_1,\ldots, R_K)\in \mc {RD}_{\mathrm{CEO}}$.
\end{new-proposition}
\begin{proof}
Let the tuple $(\Delta,R_1,\ldots, R_K)\in \mc {RI}_{\mathrm{DIB}}$ be achievable for some encoders $\phi_k$. It follows by Lemma~\ref{lem:LL_Relevance} that, by letting the decoding function $\tilde{\psi}(J_{\mc K}) = \{P_{Y^n|J_{\mc K}}(y^n|J_{\mc K})\}$, we have $\mathrm{E}[\ell_{\mathrm{log}}(Y^n,\hat{Y}^n)|J_{\mc K}]
= H(Y^n|J_{\mc K})$, and hence $(H(Y)\!-\!\Delta,R_1,\ldots, R_K)\in \mc {RD}_{\mathrm{CEO}}$. 

\noindent Conversely, assume the tuple $(D,R_1,\ldots, R_K)\in \mc {RD}_{\mathrm{CEO}}$ is achievable. It follows by Lemma~\ref{lem:LL_Relevance} that  $H(Y)-D \leq H(Y^n)- H(Y^n|J_{\mc K})=  I(Y^n;J_{\mc K})$, which implies $(\Delta,R_1,\ldots, R_K)\in \mc {RI}_{\mathrm{DIB}}$ with $\Delta = H(Y)-D$.
\end{proof}

The characterization of rate-distortion region $\mc {R}_{\mathrm{CEO}}$ has been established recently in \cite[Theorem 10]{CW14}. The proof of the theorem is completed by noting that Proposition \ref{prop:Eqreg} implies that the result in~\cite[Theorem 10]{CW14} can be applied to characterize the region $\mathcal{RI}_{\mathrm{DIB}}$, as given in Theorem~\ref{theorem-relevance-complexity-region-DM-case}.

\subsection{Proof of Proposition~\ref{proposition-parametrization-relevance-complexity-region-DM-case}}\label{appendix-proof-proposition-parametrization-relevance-complexity-region-DM-case}

 Let $\dv P^*$ be the maximizing in Equation         \eqref{eq:Dparam}. Then, 
\begin{align}
(1+s)\Delta_{s} &= (1+sK)H(Y)+sR_{s}+\mc L_{s}(\dv P^*)\\
&= (1+sK) H(Y) + sR_{ s} + \left(-H(Y|U_{\mc K}^*) - s \sum_{k=1}^K [H(Y|U^*_k) + I(X_k;U_k^*)] \right) \\
\label{eq:Ldefinition}
&= (1+sK)H(Y) + sR_{ s} +(-H(Y|U_{\mc K}^*) - s(R_s  - I(Y;U^*_{\mc K})+KH(Y)) ) \\
\label{eq:MKchain} 
&=(1+s) I(Y;U_{\mc K}^*)\\
&\leq (1+s)\Delta(R_{s}, P_{X_{\mc K},Y}), 
\label{eq:part1}
\end{align}
where Equation         \eqref{eq:Ldefinition} is due to the definition of $\mc L_s(\dv P)$ in Equation         \eqref{eq:CostF}; Equation         \eqref{eq:MKchain} holds since $
\sum_{k=1}^K [I(X_k;U_k^*) + H(Y|U_k^*)] = R_s  - I(Y;U^*_{\mc K})+KH(Y)$ using Equation         \eqref{eq:R1param}; and Equation         \eqref{eq:part1} follows by  using Equation         \eqref{eq:RelevanceSumComplexityFunction}.

Conversely, if $\dv P^*$ is the solution to the maximization in the function $\Delta(R_{\mathrm{sum}}, P_{X_{\mc K},Y})$ in Equation         \eqref{eq:RelevanceSumComplexityFunction} such that $\Delta(R_{\mathrm{sum}}, P_{X_{\mc K},Y}) = \Delta_s$, then $\Delta_s\leq I(Y;U_{\mc{K}}^*)$ and $\Delta_s\leq R -\sum_{k=1}^K I(X_k;U^*_k|Y)$ and we have, for any $s\geq 0$, that
\begin{align}
\Delta(R_{\mathrm{sum}}&,P_{X_{\mc K}, Y})  =\Delta_s \\
&\leq   \Delta_s -(\Delta_s- I(Y;U_{\mc{K}}^*)) -s\left(\Delta_s- R_{\mathrm{sum}} +\sum_{k=1}^K I(X_k;U^*_k|Y)\right) \\
&=  I(Y;U_{\mc{K}}^*)-s\Delta_s +s R_{\mathrm{sum}}  -s\sum_{k=1}^K I(X_k;U^*_k|Y) \\
\label{eq:MK}
&= H(Y) - s\Delta_s + sR_{\mathrm{sum}}-H(Y|U^*_{\mc K}) - s\sum_{k=1}^K[ I(X_k;U^*_k) +H(Y|U_k^*)]+ sKH(Y) \\
\label{eq:OptL}
&\leq  H(Y) - s\Delta_s + sR_{\mathrm{sum}}  + \mc L_{s}^* + sKH(Y) \\
\label{eq:LagrangianEq}
&= H(Y) - s\Delta_s  + sR_{\mathrm{sum}} + sKH(Y) - ((1+sK)H(Y) + sR_s -(1+s)\Delta_s) \\
&= \Delta_{s} + s(R_{\mathrm{sum}} -R_{s}),
\label{eq:LagrangianEq_2}
\end{align}
where in Equation         \eqref{eq:MK} we use  that $\sum_{k=1}^KI(X_k;U_k|Y)  = -KH(Y)+\sum_{k=1}^KI(X_k;U_k)+H(Y|U_k)$. which follows by using the Markov Chain $U_k \mkv  X_k \mkv  Y \mkv  (X_{\mc K\setminus k }, U_{\mc K\setminus k })$; Equation         \eqref{eq:OptL} follows since $\mc L^*_{s}$ is the maximum over all possible distributions $\dv P$ (possibly distinct from the $\dv P^*$ that maximizes $\Delta(R_{\mathrm{sum}}, P_{X_{\mc K},Y})$); and Equation         \eqref{eq:LagrangianEq} is due to Equation         \eqref{eq:Dparam}. Finally, Equation         \eqref{eq:LagrangianEq_2} is valid for any $R_{\mathrm{sum}}\geq 0$ and $s\geq 0$. Given $s$, and hence $(\Delta_s, R_{s})$, letting $R = R_{s}$ yields
$\Delta(R_s, P_{X_{\mc K},Y}) \leq \Delta_{ s}$. Together with Equation         \eqref{eq:part1}, this completes the proof of Proposition~\ref{proposition-parametrization-relevance-complexity-region-DM-case}.

\subsection{Proof of Lemma~\ref{lemma:QUpdate}}\label{appendix-proof-lemma:QUpdate}

Let, for a given random variable $Z$ and $z \in \mc Z$, a stochastic mapping $Q_{Y|Z}(\cdot|z)$ be given. It is easy to see that
\begin{equation}
H(Y|Z) = \mathds{E}[-\log Q_{Y|Z}(Y|Z)]-D_{\mathrm{KL}}(P_{Y|Z}\|Q_{Y|Z}).
\end{equation}
\noindent In addition,  we have
\begin{align}
I(X_k;U_k) &= H(U_k)-H(U_k|X_k)\\
&=D_{\mathrm{KL}}(P_{U_k|X_k}\|Q_{U_k}) -D_{\mathrm{KL}}(P_{U_k}\|Q_{U_k}).
\end{align}

\noindent Substituting it into Equation         \eqref{eq:CostF}, we get
\begin{align}
\mc L_s(\dv P) &= \mc L^{\mathrm{VB}}_s(\dv P, \dv Q) + D_{\mathrm{KL}}(P_{Y|U_{\mc K}}||Q_{Y|U_{\mc K}}) +s\sum_{k=1}^K (D_{\mathrm{KL}}(P_{Y|U_k}||Q_{Y|U_k})+ D_{\mathrm{KL}}(P_{U_{k}}||Q_{U_k})  ) \\
&\geq \mc L^{\mathrm{VB}}_s(\dv P, \dv Q),
\label{eq:UpperBound}
\end{align}
where Equation         \eqref{eq:UpperBound} follows by the non-negativity of relative entropy. In addition, note that the inequality  in   Equation         \eqref{eq:UpperBound} holds with equality iff $\dv Q^*$ is given by Equation         \eqref{eq:Qstarall}.

\subsection{Proof of Theorem~\ref{theorem-relevance-complexity-region-vector-Gaussian-case}}\label{appendix-proof-theorem-relevance-complexity-region-vector-Gaussian-case}

The proof of Theorem~\ref{theorem-relevance-complexity-region-vector-Gaussian-case} relies on deriving an outer bound on the relevance--complexity region, as given by Equation         \eqref{eq:ComplexityrelevanceFunction}, and showing that it is achievable with Gaussian pmfs and without time-sharing. In doing so, we use the technique of~\cite[Theorem 8]{EU14}, which relies on the de Bruijn identity and the properties of Fisher information and MMSE.

\begin{new-lemma}{\cite{DCT91,EU14}}\label{lem:FI_Ineq}
Let $(\mathbf{X,Y})$  be a pair of random vectors with pmf $p(\mathbf{x},\mathbf{y})$. We have
\begin{equation}
\log|(\pi e) \mathbf{J}^{-1}(\mathbf{X}|\mathbf{Y})|\leq h(\mathbf{X}|\mathbf{Y})\leq\log|(\pi e) \mathrm{mmse}(\mathbf{X}|\mathbf{Y})|,
\end{equation}
where the conditional Fischer information matrix is defined as
\begin{equation}
\mathbf{J}(\mathbf{X}|\mathbf{Y}) := \mathrm{E}[\nabla \log p(\mathbf{X}|\mathbf{Y})\nabla\log p(\mathbf{X}|\mathbf{Y})^\dagger]
\end{equation}
and the minimum mean square error (MMSE) matrix is 
\begin{equation}
\mathrm{mmse}(\mathbf{X}|\mathbf{Y}) := \mathrm{E}[(\dv X-\mathrm{E}[\dv X|\dv Y])(\dv X-\mathrm{E}[\dv X|\dv Y])^\dagger].
\end{equation} 
\end{new-lemma}

For $ t\in \mc{T}$ and fixed $\prod_{k=1}^{K}p(\mathbf{u}_k|\mathbf{x}_k,t)$, choose $\mathbf{\Omega}_{k,t}$, $k = 1,\ldots, K$ satisfying $\mathbf{0}\preceq\mathbf{\Omega}_{k,t}\preceq\mathbf{\Sigma}_{k}^{-1}$ such that 
\begin{align}
\mathrm{mmse}(\mathbf{Y}_k|\mathbf{X}, \mathbf{U}_{k,t},t) = \mathbf{\Sigma}_{k}-\mathbf{\Sigma}_{k}\mathbf{\Omega}_{k,t}\mathbf{\Sigma}_{k}.\label{eq:covB}
\end{align}
Note that such $\mathbf{\Omega}_{k,t}$ exists since $ \mathbf{0}\preceq\mathrm{mmse}(\mathbf{X}_k|\mathbf{Y},\mathbf{U}_{k,t},t)\preceq \mathbf{\Sigma}_{k}^{-1}$, for all $t\in \mc T$, and $k\in \mc K$. 

\noindent Using Equation         \eqref{eq:ComplexityrelevanceFunction}, we get
\begin{align}
I(\mathbf{X}_k;\mathbf{U}_k|\mathbf{Y},t)
& \geq \log|\boldsymbol\Sigma_{k}| -\log|\mathrm{mmse}(\mathbf{X}_k|\mathbf{Y},\mathbf{U}_{k,t},t) |
\nonumber\\
&
= - \log|\dv I-\mathbf{\Sigma}_{k}^{1/2}\mathbf{\Omega}_{k,t}\mathbf{\Sigma}_{k}^{1/2}|,\label{eq:firstIneq}
\end{align}
where the inequality is due to Lemma~\ref{lem:FI_Ineq}, and Equation         \eqref{eq:firstIneq} is due to Equation         \eqref{eq:covB}.

\noindent In addition, we have
\begin{align}
I(\mathbf{Y};\mathbf{U}_{S^c,t}|t)
&\leq \log|\mathbf{\Sigma}_{\dv y} |-\log|\mathbf{J}^{-1}(\mathbf{Y}|\mathbf{U}_{S^c,t},t)|\label{eq:FI_Ineq}\\
& = \log 
\left| \sum_{k\in\mathcal{S}^{c}}\mathbf{\Sigma}_{\dv y}^{1/2}\mathbf{H}_{k}^{\dagger}
\mathbf{\Omega}_{k,t}
\mathbf{H}_{k}\mathbf{\Sigma}_{\dv y}^{1/2}+\mathbf{I}\right|\label{eq:secondtIneq},
\end{align}
where Equation         \eqref{eq:FI_Ineq} is due to Lemma~\ref{lem:FI_Ineq} and Equation         \eqref{eq:secondtIneq} is due to to the following equality, which relates the MMSE matrix    in   Equation         \eqref{eq:covB} and the Fisher information,   the proof of which   follows, 
\begin{align}
\mathbf{J}(\mathbf{Y}|\mathbf{U}_{S^c,t},t) = \sum_{k\in\mathcal{S}^{c}}\mathbf{H}_{k}^{\dagger}
\mathbf{\Omega}_{k,t}
\mathbf{H}_{k}+\mathbf{\Sigma}_{\dv y}^{-1}\label{eq:Fischerequality}.
\end{align}

\noindent    To show Equation         \eqref{eq:Fischerequality}, we use de Brujin identity to relate the Fisher information with the MMSE as given in the following lemma, the proof of which can be found in~\cite{EU14}. 

\begin{new-lemma}\label{lemm:Brujin}
Let $(\mathbf{V}_1,\mathbf{V}_2)$ be a random vector with finite second moments and $\mathbf{N}\!\sim\!\mc{CN}(\dv 0, \boldsymbol\Sigma_N)$ independent of $(\mathbf{V}_1,\mathbf{V}_2)$. Then,
\begin{equation}
\mathrm{mmse}(\mathbf{V}_2|\mathbf{V}_1,\mathbf{V}_2+\mathbf{N}) = \boldsymbol\Sigma_N -\boldsymbol\Sigma_N\mathbf{J}(\mathbf{V}_2+\mathbf{N}|\mathbf{V}_1)\boldsymbol\Sigma_N.
\end{equation}
\end{new-lemma}

\noindent From the MMSE  of Gaussian random vectors~\cite{GK11},
\begin{align}
\mathbf{Y} = \mathrm{E}[\mathbf{Y}|\mathbf{X}_{\mathcal{S}^c}]+\mathbf{Z}_{\mathcal{S}^c} = \sum_{k\in \mathcal{S}^c}\mathbf{G}_{k}\mathbf{X}_{k} +\mathbf{Z}_{\mathcal{S}^c},
\end{align}
where $\mathbf{G}_{k} = \dv \Sigma_{\dv y | \dv x_{\mc{S}^c}}  \mathbf{H}^{\dagger}_{k}\mathbf{\Sigma}_{k}^{-1}$ and $\mathbf{Z}_{\mathcal{S}^c}\sim\mathcal{CN}(\mathbf{0},\dv \Sigma_{\dv y | \dv x_{\mc{S}^c}} )$,  and
\begin{align}
\dv \Sigma_{\dv  y | \dv x_{\mc{S}^c}}^{-1} =   \mathbf{\Sigma}_{\dv y}^{-1} +\sum_{k\in \mathcal{S}^c}\mathbf{H}_{k}^{\dagger}\mathbf{\Sigma}_{k}^{-1}\mathbf{H}_{k}.\label{eq:CovZ_xy}
\end{align}
Note that $\mathbf{Z}_{\mathcal{S}^c}$ is independent of $\mathbf{Y}_{\mathcal{S}^c}$ due to the orthogonality principle of the MMSE and its Gaussian distribution. Hence, it is also independent of $\mathbf{U}_{\mathcal{S}^c,q}$.

\noindent Thus, we have
\begin{align}
\label{eq:CrossTerms}
\text{mmse}\left(\sum_{k\in \mathcal{S}^c}\mathbf{G}_{k}\mathbf{X}_k\Big|\mathbf{Y}, \mathbf{U}_{\mathcal{S}^c,t},t \right)  &= \sum_{k\in \mathcal{S}^c}\mathbf{G}_{k} \text{mmse}\left(\mathbf{X}_k |\mathbf{Y}, \mathbf{U}_{\mathcal{S}^c,t},t \right)\mathbf{G}_{k}^{\dagger}\\
&= \dv \Sigma_{\dv y | \dv x_{\mc{S}^c}} \sum_{k\in \mathcal{S}^c}\mathbf{H}_{k}^{\dagger} \left(\mathbf{\Sigma}_{k}^{-1}-\mathbf{\Omega}_{k} \right)\mathbf{H}_{k}\dv \Sigma_{\dv y | \dv x_{\mc{S}^c}} 
\label{eq:CovSubs},
\end{align}
where Equation         \eqref{eq:CrossTerms} follows since the cross terms are zero due to the Markov Chain 
$(\mathbf{U}_{k,t},\mathbf{X}_k)\mkv \mathbf{Y} \mkv (\mathbf{U}_{\mathcal{K}/k,t},\mathbf{X}_{\mathcal{K}/k})$  
(see \cite[Appendix V]{EU14}); and Equation         \eqref{eq:CovSubs} follows due to Equation         \eqref{eq:covB} and $\mathbf{G}_{k}$.

\noindent Finally, we have
\begin{align}
\label{eq:LemmaBrujin}
\mathbf{J}(\mathbf{Y}|\mathbf{U}_{S^c,t},t) &=\dv \Sigma_{\dv y | \dv x_{\mc{S}^c}}^{-1} - \dv \Sigma_{\dv y | \dv x_{\mc{S}^c}}^{-1} \text{mmse} \left(\sum_{k\in \mathcal{S}^c}\mathbf{G}_{k}\mathbf{X}_{k} \Big|\mathbf{Y}, \mathbf{U}_{\mathcal{S}^c,t},t \right)  \dv \Sigma_{\dv y | \dv x_{\mc{S}^c}}^{-1} \\
\label{eq:MMSEsubs}
&=\dv \Sigma_{\dv y | \dv x_{\mc{S}^c}}^{-1} -  \sum_{k\in \mathcal{S}^c}\mathbf{H}_{k}^{\dagger} \left(\mathbf{\Sigma}_{k}^{-1}-\mathbf{\Omega}_{k,t} \right)\mathbf{H}_{k}\\
&=\mathbf{\Sigma}_{\dv y}^{-1} + \sum_{k\in \mathcal{S}^c}\mathbf{H}_{k}^{\dagger} \mathbf{\Omega}_{k,t}\mathbf{H}_{k},
\label{eq:MMSEsubs_2}
 \end{align}
where Equation         \eqref{eq:LemmaBrujin} is due to Lemma~\ref{lemm:Brujin}; Equation         \eqref{eq:MMSEsubs} is due to Equation         \eqref{eq:CovSubs}; and Equation         \eqref{eq:MMSEsubs_2} follows due to Equation         \eqref{eq:CovZ_xy}.

 Then, averaging over the time sharing random variable $T$ and letting $\bar{\mathbf{\Omega}}_k:= \sum_{t\in \mathcal{T}}p(t)\mathbf{\Omega}_{k,t}$, we get, using Equation         \eqref{eq:firstIneq},
\begin{align}
I(\mathbf{X}_k;\mathbf{U}_k|\mathbf{Y},T) 
&\geq -  \sum_{t\in \mathcal{T}}p(t) \log|\dv I-\mathbf{\Sigma}_{k}^{1/2}\mathbf{\Omega}_{k,t}\mathbf{\Sigma}_{k}^{1/2}|
\geq-\log|\dv I-\mathbf{\Sigma}_{k}^{1/2}\bar{\mathbf{\Omega}}_k\mathbf{\Sigma}_{k}^{1/2}|,\label{eq:logDetProp2}
\end{align}
where  Equation         \eqref{eq:logDetProp2} follows from the concavity of the log-det function and Jensen's inequality. 

Similarly, using Equation         \eqref{eq:secondtIneq} and Jensen's Inequality, we have
\begin{align}
I(\mathbf{Y};\mathbf{U}_{S^c}|T)
&\leq
\log\left| \sum_{k\in\mathcal{S}^{c}}\mathbf{\Sigma}_{\dv y}^{1/2}\mathbf{H}_{k}^{\dagger}
\bar{\mathbf{\Omega}}_{k}
\mathbf{H}_{k}\mathbf{\Sigma}_{\dv y}^{1/2}+\mathbf{I}\right|
\label{eq:secondtIneq_4}. 
\end{align}

The outer bound on $\mc{RI}_{\mathrm{DIB}}$ is obtained by substituting into Equation         \eqref{eq:ComplexityrelevanceFunction}, using Equations         \eqref{eq:logDetProp2} and \eqref{eq:secondtIneq_4}, noting that $\dv \Omega_k = \sum_{t\in \mathcal{T}}p(t) \dv \Omega_{k,t} \preceq\mathbf{\Sigma}_{k}^{-1}$ since $\mathbf{0} \preceq \dv \Omega_{k,t} \preceq\mathbf{\Sigma}_{k}^{-1}$, and taking the union over $\dv \Omega_k$ satisfying $\mathbf{0} \preceq \dv \Omega_k \preceq\mathbf{\Sigma}_{k}^{-1}$.
 Finally, the proof is completed by noting that the outer bound is achieved with $T= \emptyset$ and multivariate Gaussian  distributions $p^{*}(\dv u_k|\dv x_k,t) = \mc{CN}(\dv x_k,  \mathbf{\Sigma}_{k}^{1/2}(\dv \Omega_k-\dv I)\mathbf{\Sigma}_{k}^{1/2}  ) $.

\bibliographystyle{IEEEtran}
\bibliography{mybibfile}

\end{document}